\documentclass[sigconf]{acmart}

\AtBeginDocument{%
  \providecommand\BibTeX{{%
    \normalfont B\kern-0.5em{\scshape i\kern-0.25em b}\kern-0.8em\TeX}}}

\setcopyright{none}
\acmYear{2024}
\acmDOI{XXXXXXX.XXXXXXX}

\acmConference[Manuscript submitted to ACM]{Make sure to enter the correct
  conference title from your rights confirmation emai}{June 03--05,
  2018}{Woodstock, NY}
\usepackage{booktabs}
\usepackage{graphicx}

\usepackage{multicol}
\usepackage{amsmath}
\usepackage{dsfont}
\usepackage{algorithm}
\usepackage{algpseudocode}
\usepackage{newtxmath}
\usepackage{adjustbox}
\DeclareUnicodeCharacter{0229}{\c{e}}

\usepackage{multirow} 
\usepackage{array} 
\usepackage{colortbl}
\usepackage{caption}
\usepackage{subcaption}

\usepackage[utf8]{inputenc}
\usepackage{enumitem}

\begin{document}

\title{Biodegradable Interactive Materials}

\author{Zhihan Zhang}
\orcid{0000-0001-7394-5409}
\affiliation{%
  \institution{University of Washington}
  \city{}
  \state{}
  \country{}
}
\email{zzhihan@cs.washington.edu}

\author{Mallory Parker}
\affiliation{%
  \institution{University of Washington}
  \city{}
  \state{}
  \country{}
}
\email{parkerme@uw.edu}

\author{Kuotian Liao}
\affiliation{%
  \institution{University of Washington}
  \city{}
  \state{}
  \country{}
}
\email{timliao@uw.edu}

\author{Jerry Cao}
\affiliation{%
  \institution{University of Washington}
  \city{}
  \state{}
  \country{}
}
\email{jcao22@uw.edu}

\author{Anandghan Waghmare}
\affiliation{%
  \institution{University of Washington}
  \city{}
  \state{}
  \country{}
}
\email{anandw@cs.washington.edu}

\author{Joseph Breda}
\affiliation{%
  \institution{University of Washington}
  \city{}
  \state{}
  \country{}
}
\email{joebreda@cs.washington.edu}

\author{Chris Matsumura}
\affiliation{%
  \institution{University of Washington}
  \city{}
  \state{}
  \country{}
}
\email{matsuc@uw.edu}

\author{Serena Eley}
\affiliation{%
  \institution{University of Washington}
  \city{}
  \state{}
  \country{}
}
\email{serename@uw.edu}

\author{Eleftheria Roumeli}
\affiliation{%
  \institution{University of Washington}
  \city{}
  \state{}
  \country{}
}
\email{eroumeli@uw.edu}

\author{Shwetak Patel}
\orcid{0000-0002-6300-4389}
\affiliation{%
  \institution{University of Washington}
  \city{Seattle}
  \state{WA}
  \country{USA}
}
\email{shwetak@cs.washington.edu}

\author{Vikram Iyer}
\orcid{0000-0002-3025-7953}
\affiliation{%
  \institution{University of Washington}
  \city{Seattle}
  \state{WA}
  \country{USA}
}
\email{vsiyer@uw.edu}

\renewcommand{\shortauthors}{Zhang, et al.}

\begin{abstract}
  The sense of touch is fundamental to how we interact with the physical and digital world. Conventional interactive surfaces and tactile interfaces use electronic sensors embedded into objects, however this approach poses serious challenges both for environmental sustainability and a future of truly ubiquitous interaction systems where information is encoded into everyday objects. In this work, we present Biodegradable Interactive Materials: backyard-compostable interactive interfaces that leverage information encoded in material properties. Inspired by natural systems, we propose an architecture that programmatically encodes multidimensional information into materials themselves and combines them with wearable devices that extend human senses to perceive the embedded data. We combine unrefined biological matter from plants and algae like chlorella with natural minerals like graphite and magnetite to produce materials with varying electrical, magnetic, and surface properties. We perform in-depth analysis using physics models, computational simulations, and real-world experiments to characterize their information density and develop decoding methods. Our passive, chip-less materials can robustly encode 12 bits of information, equivalent to 4096 unique classes. We further develop wearable device prototypes that can decode this information during touch interactions using off-the-shelf sensors. We demonstrate sample applications such as customized buttons, tactile maps, and interactive surfaces. We further demonstrate the natural degradation of these interactive materials in degrade outdoors within 21 days and perform a comparative environmental analysis of the benefits of this approach.

\end{abstract}



\begin{teaserfigure}
  \includegraphics[width=\textwidth]{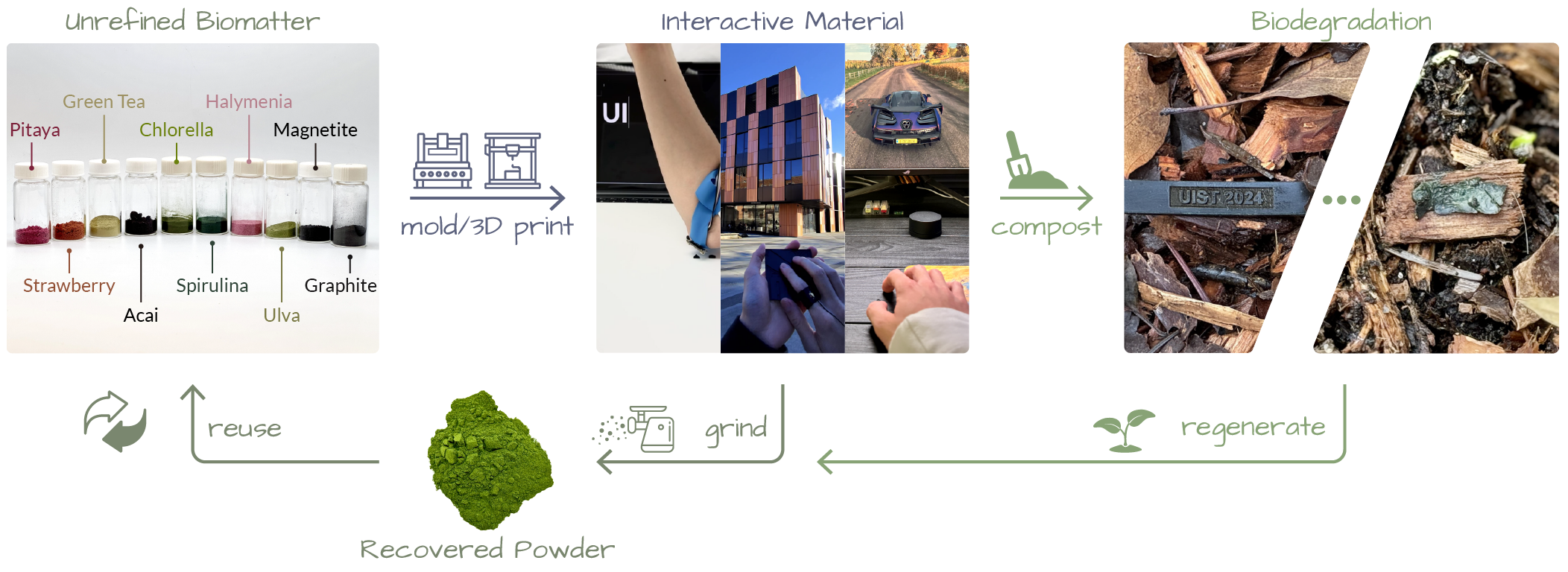}
  \caption{Circular life cycle of Biodegradable Interactive Materials. These interactive materials are created from unrefined biological matter and fabricated through simple and scalable techniques. The passive, chip-less tactile interfaces enable ubiquitous surface interactions. At the end-of-life, these materials can either be ground into powder for immediate reuse or composted into backyard soil for natural biodegradation. The raw biomatter will be regenerated through the natural biological cycle.}
  \Description{}
  \label{fig:teaser}
\end{teaserfigure}

\received{20 February 2007}
\received[revised]{12 March 2009}
\received[accepted]{5 June 2009}

\maketitle

\section{Introduction}







The sense of touch is among the most basic human senses and is fundamental to how we interact with the physical and digital world~\cite{meltzoff_infant_2018}. From buttons and knobs to touchscreens, interactive surfaces, and ID tags, tactile interfaces have become natural and ubiquitous means for our electronic devices to sense how we interact with them. The conventional approach to designing interactive surfaces and tactile interfaces is to use electronic sensors embedded into objects that transform mechanical motion or human-generated electric fields into changes in resistance, capacitance, or other properties that can be digitized \cite{davis_tangiblecircuits_2020, li_editing_2019, melfi_understanding_2020, zhang_interactiles_2018, sato_touche_2012, holloway_accessible_2018, cheng_swellsense_2023}. This approach however poses serious challenges both for environmental sustainability and a future of truly ubiquitous interaction systems where information is encoded into everyday objects all around us~\cite{abowd_internet_2020}. In addition to the high carbon cost of fabricating semiconductors~\cite{gupta_act_2022, zhang_deltalca_2024} and energy consumed during use and data processing in cloud-connected systems~\cite{anderson_treehouse_2022}, electronics contain hazardous materials that can damage the environment and human health at end-of-life disposal~\cite{ogunseitan_biobased_2022}. Beyond environmental costs, the physical size of bulky sensing circuits and their need for power prevent widespread integration into everyday objects.

How can we develop an alternative paradigm for electronic systems to perceive tactile interactions? If we look to nature, we see numerous examples of how animals have adapted specialized perception systems to interact with natural materials in the world around them. For example, bees can sense electric fields in flowers, turtles and other animals can detect magnetic fields, and animals like spiders can detect minute surface vibrations. These animals utilize their specialized sensory organs to perceive the world, rather than instrumenting the world with external sensors that then provide feedback. Inspired by natural systems, we propose an architecture for tactile interactive systems that programmatically encodes multi-dimensional information into materials themselves and combines them with wearable devices that extend human senses to perceive the embedded data.

In this work, we develop \textit{Biodegradable Interactive Materials} that can store custom embedded data and systems to read this data using wearable devices. \textcolor{blue}{Figure~\ref{fig:teaser}} shows the end-to-end lifecycle of these materials made of unrefined biomatter from plants like strawberries and algae like chlorella. We combine these materials with natural minerals likegraphite and magnetite to modulate their electrical and magnetic properties to encode information. We develop digital fabrication techniques like 3D printing and compression molding to encode a third dimension of information using surface textures and energy. At the end-of-life, these objects can be ground up for recycling or biodegraded in backyard soil without any specialized facilities. Additionally, these materials are completely passive which enables truly widespread, ubiquitous deployments by attaching them like buttons or tags to existing objects or even creating new large-scale objects like tabletops and surfaces with embedded information.

To create end-to-end interactive systems with these novel materials, we develop wearable prototypes to decode this information using off-the-shelf hardware such as bio-impedance sensors, magnetometers, and microphones. We perform detailed theoretical, simulated, and experimental analyses to determine the information density limits for each material property. Even with the resolutions of these common sensors across multiple finger positions and angles, our three dimensions allow us to robustly encode a total of 12 bits representing 4096 distinct classes. We use these primitives to demonstrate prototypes of end-to-end interaction systems such as customizable label inputs, tactile maps in which touch triggers audio feedback for Blind or Low-Vision (BLV) individuals, and programmable surfaces for interactive games.

We summarize our key contributions below:
1) We use unrefined biological matter to create the first passive, chip-less, tactile interfaces to enable sustainable, ubiquitous interactive systems. To achieve this we develop an end-to-end fabrication pipeline that combines biomatter with conductive and magnetic materials to produce textured objects with embedded information.

2) We develop wearable systems that can decode the data using common, off-the-shelf sensors such as bio-impedance sensors, magnetometers, and microphones. We develop highly detailed physics models, and computational simulations and perform experiments to show that by using these 3 dimensions, we can robustly embed 12 bits of data representing 4096 classes.

3) We conduct extensive experiments with our biodegradable interactive materials and demonstrate three end-to-end applications. We further perform a comparative environmental impact analysis to emphasize the environmental benefits of our approach, and outdoor experiments demonstrating our prototypes can biodegrade naturally in backyard soil within 21 days.

\section{Related Work}
This interdisciplinary work combines classic physics, sustainable materials, fabrications, sensing, and ubiquitous computing to create biodegradable \textit{interactive materials} (IM). We present a survey of works on related topics below. 

\subsection{Internet of Materials}
A vision talk at UIST 2018 by Gregory Abowd proposed going beyond the Internet of Things (IoT) to create the Internet of Materials (IoM) \cite{abowd_internet_2020}. This vision proposes that the materials themselves that make up everyday objects have computing "woven" into them and behave as connected computational entities. Since then only a handful of works have fully met the proposed criteria for ``self-sustainable computational materials'' \cite{arora_saturn_2018,arora_mars_2021}. A larger body of work has instead focused on integrating energy harvesting into traditional battery-powered IoT designs using solar cells \cite{zhang_optosense_2020, zhang_flexible_2022, waghmare_ubiquitouch_2020, de_winkel_battery-free_2020}, piezoelectric generators \cite{lu_battery-less_2020, jang_underwater_2019}, or NFC \cite{kim_remote_2023}.
The gap between the current state of technology and the envisioned future of IoM underscores the significant challenges of creating computational objects that simultaneously address self-sustainability in terms of power and life cycle, as well as manufacturing scalability while maintaining practical form factors of everyday physical objects. In this work, we take a different approach in which we develop mechanisms to encode static data into materials, and combine these with wearable sensors to create passive, chip-less tactile interfaces.

\subsection{Internet of Bioinspired Things}
\begin{figure}[t]
    \centering
    \includegraphics[width=\linewidth]{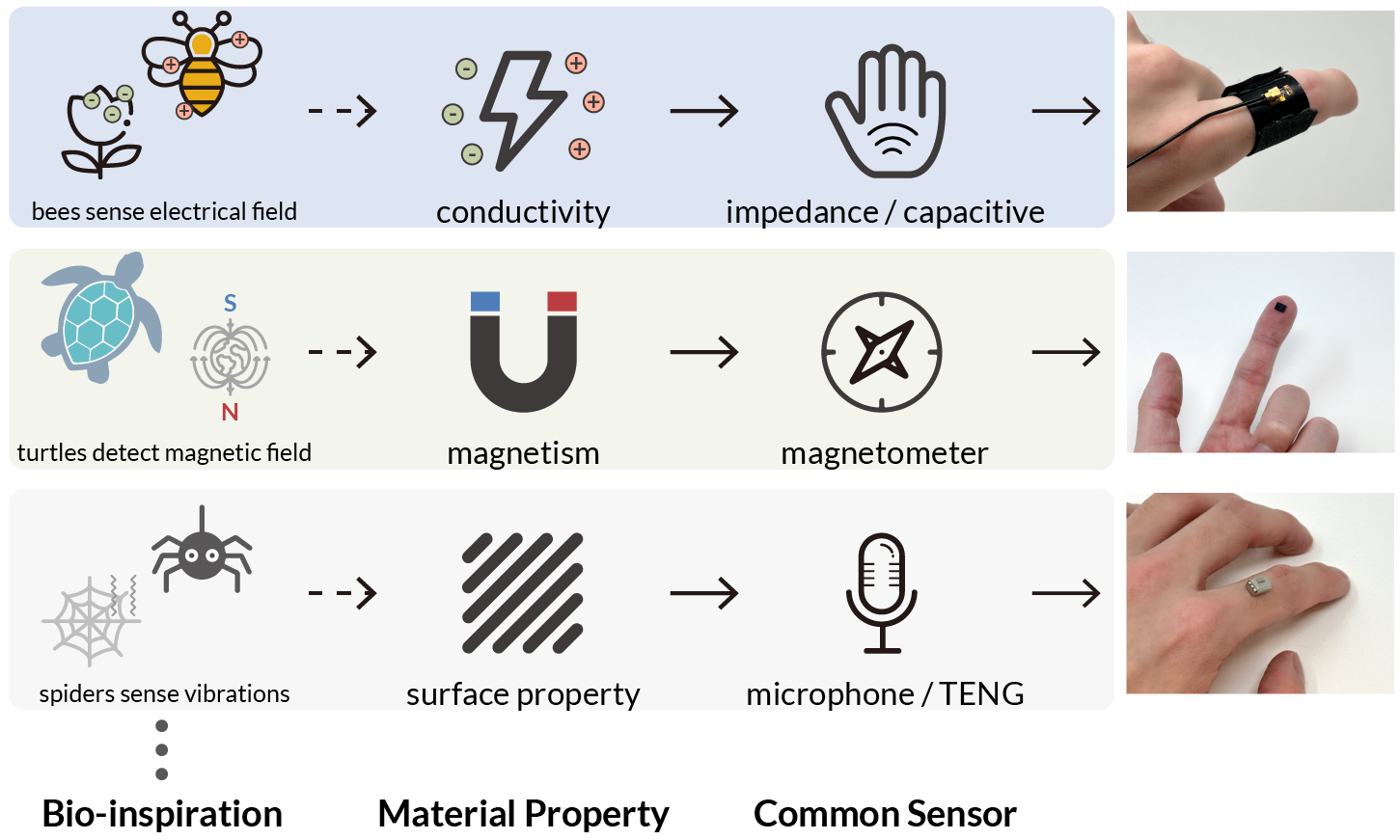}
    \caption{Bio-inspired information encoding through natural material properties. 1) An off-of-shelf bio-impedance sensor can measure impedance changes when touching a conductive surface, akin to bees sensing a flower's electric field. 2) Magnetometers can measure magnetic field strength, akin to turtles sensing intensity and inclination of Earth's magnetic field for navigation. 3) Microphones can detect the vibrations of human motion on a tangible surface, akin to spiders can sense vibrations when prey touch their webs.}
    \label{fig:overview}
\end{figure}

Recent work has explored bioinspired approaches to developing new capabilities for IoT and cyber-physical systems \cite{iyer_wireless_2020, iyer_wind_2022, johnson_solar-powered_2023}. These works focus primarily on environmental sensing and combine insights into biological systems, such as the shape of dandelion seeds, with advances in wireless systems (e.g., backscatter communication). We apply a similar design approach to create interactive materials by leveraging insights into how nature’s entities, from honeybees to sea turtles and spiders, interact with their environment through innate mechanisms. These natural phenomena offer rich ideas for our design of interactive materials (\textcolor{blue}{Figure~\ref{fig:overview}}).


\subsection{Sustainable Computing in HCI}
The urgency of sustainability resonates strongly within the HCI community which has seen a growing body of work on this topic. This includes the development of bio-based \cite{vasquez_myco-accessories_2019, koelle_prototyping_2022, nicolae_biohybrid_2023} and recyclable materials \cite{zhang_recyclable_2023} for circuits, aiming to extend the lifecycle of electronic components while circumventing the environmental pitfalls associated with processing the raw material of traditional bioplastics. Despite these advances, such biomaterials are often come with trade-offs of suboptimal mechanical properties and challenging processability issues, rendering them impractical for making everyday objects. Additionally, while recent works have proposed sustainable designs using water-soluble plastics such as polyvinyl alcohol (PVA) \cite{cheng_functional_2023}, these materials do not fully degrade in water leaving contaminants that are beyond most wastewater systems \cite{rolsky_degradation_2021}. In contrast, we seek to design materials that are both highly functional and can degrade fully in soil without harmful impacts. 


\subsection{Sensing for Interactive Surface}
Traditional sensing for surface interactions (e.g., touch detection) integrates sensors within physical interfaces, including capacitive \cite{sato_touche_2012, holloway_accessible_2018, cheng_swellsense_2023} and vibration sensors \cite{arora_saturn_2018, ding_surface_2023}. Beyond these, alternative strategies have explored the use of vision-based systems, mounting depth \cite{wilson_using_2010, xiao_direct_2016}, or thermal cameras \cite{larson_heatwave_2011} above a given surface. These vision-based approaches, while powerful, often involve high setup costs and operational computational overheads for processing video data. Such requirements pose substantial barriers to the ubiquitous deployment of interactive surfaces into everyday environments, limiting their practicality. Additionally, vision requires line of sight and does not enable the programmable embedding of information. Prior work has also explored embedding information in 3D printed objects using magnetic material~\cite{iyer_3d_2017}, however are not biodegradable, do not develop wearable sensors, and require specific motions over large areas of the object to decode data.

\section{Designing Biodegradable Interactive Materials}
This section presents an overview of Biodegradable Interactive Materials, covering our selection of material properties for information encoding, a curated list of sustainable materials, and the considerations underpinning our sustainable sensing design. For each material property explored, we aim to decode the information leveraging common, off-the-shelf sensors (e.g., impedance sensors, magnetometers, microphones) to make this technology accessible to the broader HCI community, rather than developing custom sensors which typically incur prohibitive fabrication costs and times. 

\subsection{Bioinspired Information Encoding}
The key primitive required for creating biodegradable interactive materials is a means to encode information in an object. This allows a wearable sensor to detect interactions such as a specific button press and map it to the appropriate output. To identify properties of biodegradable materials that could be sensed by wearables, we take inspiration from the natural world by looking to specialized senses evolved by different animals (\textcolor{blue}{Figure~\ref{fig:overview}}).

\subsubsection{Electrical Properties}
Certain animals, like honeybees, produce sensory hairs to perceive the electrical properties of objects. Honeybees accumulate positive charges while flying, and flowers gain negative charges through electrostatic induction. Honeybees induce an electric field change on visited flowers. Sensing these fields enables bees to determine the likelihood of nectar availability before landing \cite{clarke_bee_2017}. These observations suggest electrical properties such as conductivity as a medium to encode information in natural materials. For example, embedding natural conductors such as graphite into a material will modify its impedance and how it responds to electrical signals.

\subsubsection{Magnetic properties}
Many species of animals have evolved to perceive magnetic fields. For example, sea turtles and migratory birds sense Earth's magnetic for navigation \cite{kishkinev_navigation_2021}. While the exact mechanisms behind biological magnetoreception are an active area of research, animals ranging from bacteria to birds contain magnetite crystals~\cite{blakemore_magnetotactic_1975}. Inspired by this, we incorporate magnetite into our materials which can be easily sensed using commercial magnetometers. 

\subsubsection{Surface properties}
In addition to electromagnetic senses, many creatures have evolved remarkably sensitive mechanoreceptors sensitive to properties and vibrations on a surface. For example, spiders, through their slit sensillae (a form of mechanoreceptors), are capable of detecting minute vibrations and surface irregularities for discerning varied terrains for web placement and detecting prey caught in their webs. We can extend this idea of vibration sensing to measure the response of human motion such as a simple swipe gesture on surfaces with varying textures. 

We summarize these material properties and their associated sensing technologies in \textcolor{blue}{Figure~\ref{fig:overview}}. Next, we describe our fabrication and environmental impact, followed by a series of sections covering the relevant theory, simulation, and implementation for encoding information with each material property.


\subsection{Fabrication and Disposal}

\subsubsection{Sustainable Material Selection.} 
Addressing the limitations associated with traditional biomaterials, such as inferior mechanical properties and challenging processability, we leverage recently demonstrated backyard-compostable bioplastics from raw microalgae \cite{campbell_progress_2023} as our base material, and add conductive and/or magnetic materials and surface treatments to encode information. This material allows for a simple and scalable fabrication methodology such as heat pressing \cite{iyer_fabricating_2023, liao_effects_2023} and 3D printing \cite{fredricks_spirulina-based_2021} and can accommodate a wide variety of unrefined biomatter as shown in Fig~\ref{fig:teaser}. These techniques yield bioplastics with mechanical properties comparable to those of conventional plastic filaments. As a proof of concept, we select chlorella (a type of edible microalgae; powder, Nuts, USA) as the base material, along with graphite (powder, Sigma Aldrich, USA) and magnetite Fe3O4 (a natural mineral; powder, Alpha Chemicals, USA) to tune the material properties. We vary the weight percentage of graphite and magnetite to encode user-defined information.

\begin{figure}[t]
    \centering
    \includegraphics[width=\linewidth]{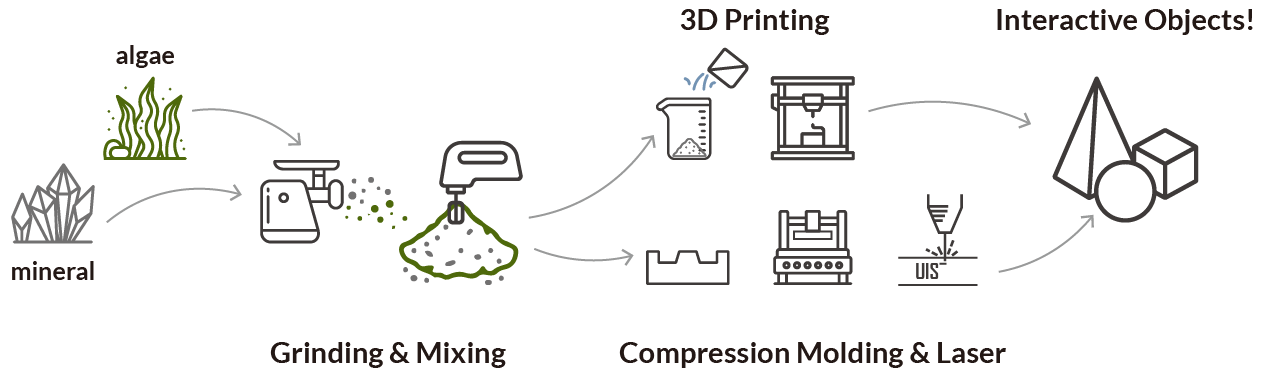}
        \caption{Interactive material fabrication. We mix unrefined biomatter powder with varying weight percentages of minerals like graphite and magnetite, then add water for 3D printing or direct compression molding and laser cutting to create interactive everyday objects with surface features.}
    \label{fig:fabrication}
\end{figure}

\subsubsection{Fabrication Process.} To fabricate our biodegradable IMs, we first mix graphite or/and magnetite powder as filler in the chlorella matrix using a VWR Analog Vortex Mixer at varying weight percentages (0 wt\% for pure chlorella, and incrementally up to 28 wt\%). For the heat pressing process, the powder mixtures are compression molded in custom stainless steel molds on a TMAX-RT24D hot press at 15 MPa and 135 °C for 5 minutes. Next, a laser micromachining system (LPKF ProtoLaser U4) is used to pattern surface features. We tune the laser parameters carefully to minimize any heat-induced carbonization to the material surface. For the 3D printing process, we mix the powder mixtures with deionized water at a weight ratio 1:1.2 (powder mixtures: water), then use a 3D Bioprinter (Direct-Ink-Writing; BioBot Basic, Advanced Solutions, USA) to print the powder mixtures, following by drying \cite{fredricks_spirulina-based_2021}. We note that while both fabrication methods prove effective, the heat pressing technique generally yields objects with superior mechanical integrity, attributable to the bonding effect induced by the combined application of heat and pressure.

\subsubsection{Biodegradation.}
\begin{figure}[h]
    \centering
    \includegraphics[width=\linewidth]{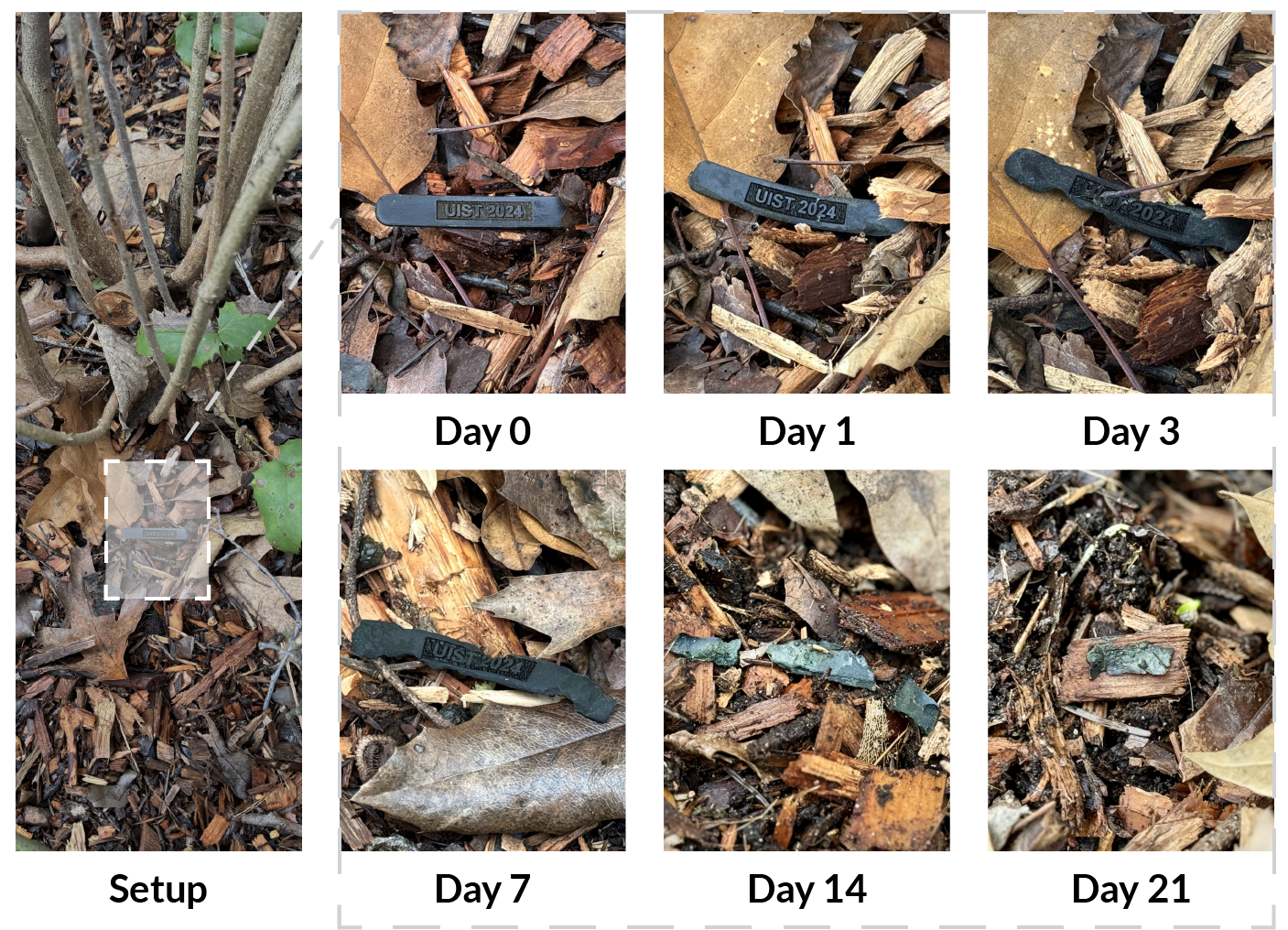}
    \caption{Biodegradation in soil. Photographs showing the biodegradation process of Biodegradable Interactive Materials disposed directly on backyard soil, the specimen is observably decomposed on Day 21. Note that on Day 7, a (white) centipede was eating the specimen (bottom-left).}
    \label{fig:biodegradation}
\end{figure}

To demonstrate the backyard degradability of our IMs, we compost an IM (90 wt\% chlorella and 10 wt\% graphite) with laser-cut text "UIST 2024" in a yard located in the United States
during the winter of 2024, spanning a duration of 21 days. The specimen was "disposed" directly on the soil surface, eschewing burial, to simulate a common disposal scenario. Daily weather conditions are recorded and detailed in \textcolor{blue}{Table A\ref{tab:biodegradation_daily_weather_data}} using the data from Weather Underground\footnote{\noindent\href{https://www.wunderground.com/}{https://www.wunderground.com/history/}}.
Photographic documentation captured the progressive decomposition of the IM over the trial period with discernable signs of decomposition, as shown in \textcolor{blue}{Figure \ref{fig:biodegradation}}. We note that many weather factors vary the degradation rates, for example, we found that dramatic shifts in temperature and humidity (e.g., a sunny day with a high UV index following a rainy day; day 7 to 8 and day 13 to 14) or persistent rain (day 17 to 21) would accelerate the degradation process.

We imagine the future of sustainable materials extracted from renewable biological matter and processed through eco-friendly techniques to minimize environmental burdens.


\subsection{Evaluating Environmental Impact}

\begin{table*}[t]
  \centering
  \includegraphics[width=\linewidth]{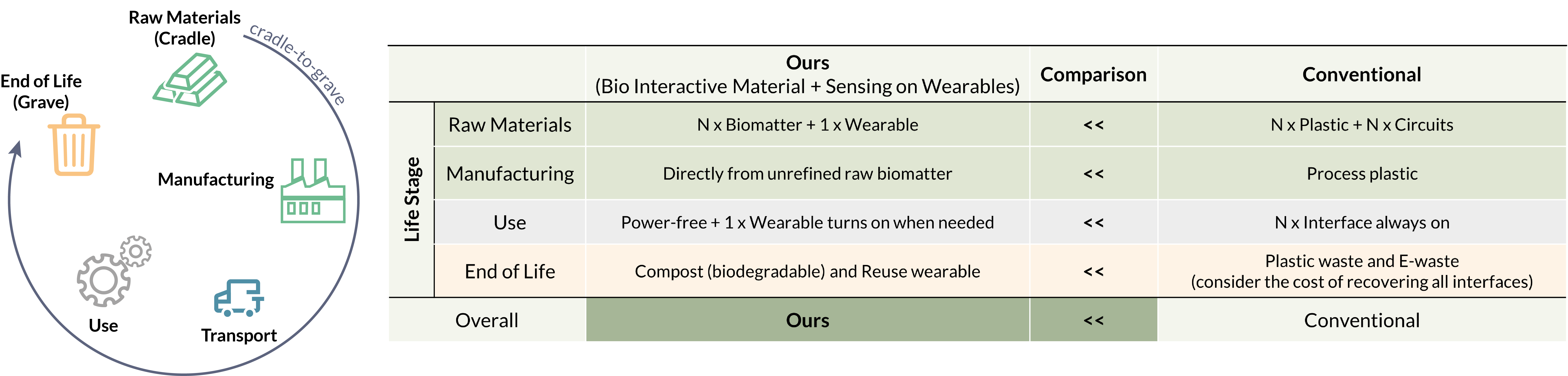}
    \caption{Environmental impact comparison of our method (biodegradable interactive material + sensing on wearables) against conventional approaches of creating interactive surfaces. Our method has significant environmental benefit potential in the future world of ubiquitous interactive surfaces.}
    \label{tab:LCAComparison_Main}
\end{table*}

The experiments above show that our biodegradable materials naturally decompose, however creating an end-to-end interaction system still requires electronics. Our approach of decoupling sensing electronics onto a wearable device offers numerous sustainable advantages over conventional design approaches, which embed electronic sensors into the interaction interface: 1) it centralizes the sensing mechanism, reducing the number of electronics per user to a single wearable device. This reduction is critical for minimizing environmental waste when creating truly scalable and ubiquitous interaction devices. 2) the personal nature of wearables facilitates better end-of-life treatment and power management, including the option to activate the device only when needed, thereby minimizing unnecessary power consumption. 



We perform a cradle-to-grave environmental impact comparison taking a similar approach to recently developed comparative life-cycle assessment (LCA) methods\cite{zhang_deltalca_2024} in \textcolor{blue}{Table~\ref{tab:LCAComparison_Main}} (see \textcolor{blue}{Tables A\ref{tab:detailed_LCA}} for detailed comparisons of scenarios involving single and multiple interactive interfaces). We juxtapose our biodegradable interactive materials with the wearable-based sensing strategy against traditional methods involving 3D-printed physical interfaces equipped with embedded capacitive sensing circuits. This analysis reveals multiple sustainability benefits. First, our method uses unrefined biomatter --- this is a wasteless transformation, all biomass is used without separation and generating waste streams, also avoiding costly and polluting resource extraction. Furthermore. Second, the one-to-many architecture of a wearable sensor reduces the number of electronic circuits required and their power consumption during their use phase. Third, at the end of life, our interactive materials will naturally degrade without environmental harm.

\section{Electrical Encoding and Sensing}

Building on our inspiration from the electroreception of honeybees, in this section, we aim to encode information into materials through conductivity and extend human sensory capabilities by leveraging the recent development of wearable bio-impedance sensing technology \cite{waghmare_z-ring_2023, zhang_advancing_2016, jung_wrist-wearable_2021}. Our approach aims to replicate, in human users, a bee-like ability to discern and interact with the electrical properties of their surroundings. We describe the theoretical foundations for this approach, followed by simulations and experimental data.

\subsection{Theory}

We encode information in a material's electrical conductivity $\sigma$ and decode the data using wearable bio-impedance sensors. To explore the information density limits and decoding strategies of this approach, we develop a simplified circuit model by combining classical electrical theory and bioelectrical impedance analysis to model a user's finger touching our biodegradable conductive materials.

We assume that the user's finger and the external surface form a series connection upon contact. The system's total impedance, $Z_{total}$, is the sum of the finger's bio-impedance $Z_{bio}$, the contact electrical impedance $Z_{contact}$, and the inherent material impedance $Z_{material}$:
\begin{equation}
   Z_{total} = Z_{bio} + Z_{contact} + Z_{material}
\label{eq:Z_total}
\end{equation}

Bio-impedance, a measure of biological tissues' opposition to electric current, varies with tissue type and state, such as hydration. In our case, we neutralize the variable of the user's inherent bio-impedance by only focusing on the change in impedance $\Delta Z$ upon contact with an external surface, i.e., the add-in contact and IM impedance:
\begin{align}
    \Delta Z &= Z_{total,new} - Z_{total,initial} \nonumber \\
    \Rightarrow \Delta Z &= (Z_{bio} + Z_{contact} + Z_{material}) - Z_{bio} \nonumber \\
    &= Z_{contact} + Z_{material}
\label{eq:delta-Z}
\end{align}

For the contact impedance, we use the Maxwell spreading resistance formula to constrain the electron scattering mechanism, and further refine through the Hertzian contact stress model to express the effective contact radius $a$ of the finger pad and IM. We derive the contact impedance, $Z_{contact}$, as a function of material resistivity (refer to \textcolor{blue}{Section \ref{Detailed_ContactImpedance}} for detailed derivation):

\begin{equation}
    Z_{contact} = \frac{11 \Omega \text{m}+\rho_{material}}{4a}, a \in [0.84 mm, 2.40 mm]
\label{eq:Wexler_final}
\end{equation}
where $\rho_{material}$ is the resistivity of IM.

We further derive the intrinsic impedance of IM, $Z_{material}$ in higher frequency alternating current (AC) conditions for sensing with capacitance and bio-impedance sensors (detailed derivation in \textcolor{blue}{Section \ref{Detailed_MaterialImpedance}}):

\begin{equation}
    Z_{material} = \rho \frac{t}{A} - j\frac{t}{\omega \epsilon A} = \frac{t}{A} (\rho - j\frac{1}{\omega \epsilon})
\label{eq:Z_material_final}
\end{equation}
where $j$ is the imaginary unit, $\omega$ is the angular frequency $2\pi f$; $\epsilon$, $A$, $t$ are the permittivity, surface area and thickness of IM, respectively. 

For the minimal dimensions of interactive material pixel, we adhere to Apple’s Accessibility Human Interface Guidelines \cite{apple_accessibility_nodate}, which suggest that a hit target should measure at least 44$\times$44 points, approximately 15$\times$15 mm. We envision our interactive materials being applied to the surface of everyday objects rather than filling the entire solid object. Consequently, they will be thin, such that $t\ll A$. Therefore, the impedance of the interactive material is negligible in our final calculation except in a low-frequency region (e.g., $f<1kHz$), where the term $(\rho - \frac{1}{\omega \epsilon})$ is dominated by the permittivity. We therefore simplify the impedance change Eq.\ref{eq:delta-Z} as: 
\begin{align}
     \Delta Z &= Z_{contact} + Z_{material} \approx Z_{contact} \nonumber \\
     &= \frac{11 \Omega \text{m}+\rho_{material}}{4a}, a \in [0.84 mm, 2.40 mm], f > 1MHz
\label{delta-Z-final}
\end{align}

This equation allows us to calculate the maximum and minimum conductivity values we can achieve in our IMs. Next, we seek to discretize this range into states that can be mapped to bits. To optimize the signal-to-noise ratio (SNR) and read margin between each possible conductivity states, we consider the variance in touch force and user-specific biological tissue properties as background noise or the uncertainty region. We set the resistivity bounds of $\rho_{mateiral}$ from $1.86\times10^8 \Omega \text{m}$ (characterized resistivity of chlorella, see resistivity measurement in \textcolor{blue}{Section \ref{conductivity_implementation}}) to $\rho_{finger}$ $11 \Omega \text{m}$. This delineation is predicated on the principle that if the interactive material exhibits superior conductivity like copper (i.e., $\rho_{material} \ll \rho_{finger}$), then the contact resistance part will be dominated by $\rho_{finger}$. Under such circumstances, the dimension—specifically, the area ($A$) of the interactive material—becomes a tunable parameter for encoding information, however this is a different material property and will not be discussed here. 

In conclusion, by tuning the conductivity of the material, it becomes feasible to encode information within a spectrum of approximately 32 discrete states, equivalent to 5 bits of data, as expressed by the following equation:

\begin{equation}
    \log_{1.68}{(1.86\times10^8 \Omega \text{m} \div 11 \Omega \text{m})} \approx 32 \text{ states} \Leftrightarrow \log_{2}{32} = 5 \text{ bits}
    \label{eq:conductivity_result}
\end{equation}

Through this theoretical exposition, our objective is not only to elucidate the boundaries of information encoding within materials via conductivity but also to outline the mechanism by which a wearable impedance sensor could decode this information.

\begin{figure*}[h]
    \centering
    \includegraphics[width=\textwidth]{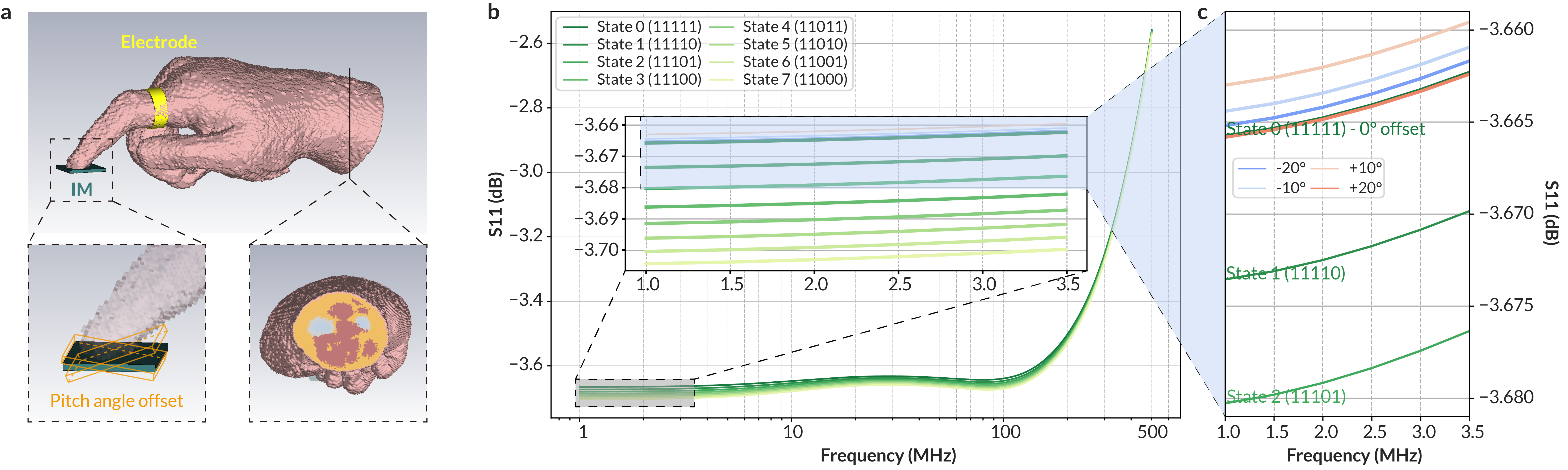}
    \caption{Electrical simulation. a, Simulation setup of conductive interactive material and bio hand model in CST Studio Suite, as well as pitch angle offset setup and the cross-section of bio hand model showing various body tissues. b, Simulated S11 results, showing the amount of signal reflected caused by impedance mismatch resulting from contact with external surfaces with corresponding conductive levels. c, Simulated S11 results respond to varying pitch angle offsets.}
    \label{fig:conductivity_simulation}
\end{figure*}

\subsection{Simulation}


We evaluate our hypotheses and conductivity model using Finite Integration Technique (FIT) simulations. We conduct simulations using the Simulia CST Studio Suite, leveraging its 3D electromagnetic (EM) analysis capability. We select a frequency range of 1 MHz to 500 MHz for our simulations, drawing on recent findings highlighting a significant attenuation of specific absorption rate (SAR) dynamics within human tissue beyond 500 MHz \cite{waghmare_z-ring_2023}.

We constructed a bio hand model representative of a 38-year-old male, 180 cm in height and 103 kg in weight, utilizing the CST Bio Models Library and the US National Library of Medicine's Visible Human Project. Our model incorporates 31 distinct tissue types at a resolution of $1\times1\times1 \text{mm}$, and isolates the right hand from the rest of the body tissues to reduce the size of mesh volume and computation complexity. The bio-impedance measurement point is set at the proximal phalange (the common position to wear a ring-form-factor device) of the index finger (the common finger to touch an external object). 

The interactive materials are created with dimensions of $20\times20\times5 \text{mm}$, balancing the requirements for minimal accessible touch size and interactive material 'pixel' density. Within this framework, conductivity emerges as the sole variable, with all other material properties, such as loss tangent, held constant to isolate the effect of other contributions. The simulation models the contact interface between the fingerpad and interactive materials, following our calculation in Eq.\ref{eq:radius_number} of contact radius and pressure distribution. The final model setup is illustrated in \textcolor{blue}{Figure~\ref{fig:conductivity_simulation}a}.

To explore the conductivity spectrum of the interactive materials, we conduct a logarithmic parameter sweep from $5.37\times10^{-9}$ S/m (state 0; characterized conductivity of chlorella, see conductivity measurement in \textcolor{blue}{Section \ref{conductivity_implementation}}) to $6.87\times10^{-7}$ S/m (state 7) with a total of 8 samples. This interval selection is predicated on the minimal read margin calculated in Eq.\ref{eq:conductivity_result} to achieve a good SNR, ensuring discernible differentiation between each conductive state. Each parameter value required approximately 100 minutes for simulation on a 14-core CPU and a single RTX 4070Ti GPU. The measurements of S11, also known as the reflection coefficient, are used to determine the impedance. S11 is a metric that specifies the amount of a wave reflected by the impedance mismatch in the transmission medium, defined as the ratio of the reflected wave’s amplitude to the incident wave’s amplitude. The simulation results reveal macroscopic distinctions across 8 conductivity levels as shown in \textcolor{blue}{Figure~\ref{fig:conductivity_simulation}b}, corroborating our theoretical predictions.

To enable robust decoding, we further simulate the influence of finger orientation on touch interaction. We perform another linear parameter sweep of touch pitch angles from -20 degrees to +20 degrees with a total of 5 samples onto the existing model. The S11 simulation results, along with the nearest conductivity states as references, are shown in \textcolor{blue}{Figure~\ref{fig:conductivity_simulation}c}, revealing the robustness of conductivity detection across a range of touch orientations. The aggregation of these 3D EM simulation results validates our theoretical constructs.


\begin{figure*}[t]
    \centering
    \includegraphics[width=\textwidth]{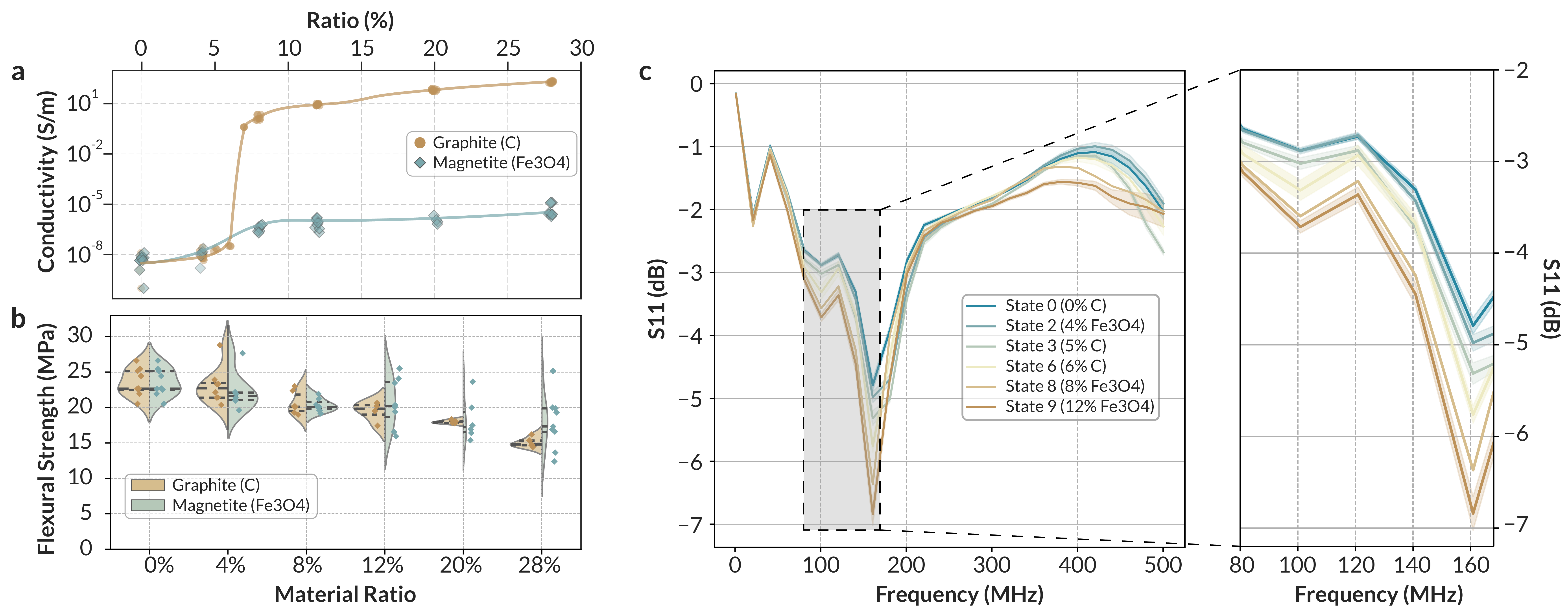}
    \caption{Electrical implementation. a,b, Characterized conductivity (a) and flexural strength (b) of Interactive Materials with different proportions of graphite and magnetite, data is presented as mean (SD) of 3 (a) and 9 (b) specimens in 1000 (a) and 1 (b) measurements of each proportion. c, Measurements from a wearable bio-impedance sensor when user's finger is in contact with external surfaces of corresponding conductive states ($N=2\times5\times2\times30=600$ frequency sweeps per state, shaded region indicates SD).}
    \label{fig:conductivity_implementation}
\end{figure*}

\subsection{Implementation} \label{conductivity_implementation}
The simulation and theoretical analysis above show the density of information we can encode with conductivity. Next we fabricate the materials and perform empirical measurements. This phase entails the fabrication of interactive materials with varied conductivities, alongside the development of a wearable sensor system capable of decoding this conductivity. We aim to validate the feasibility of our approach in encoding information within materials through their inherent conductivity properties in practice.

We fabricate interactive material samples with different conductivities by mixing chlorella with varying proportions (0\%--28\%) of graphite or magnetite Fe3O4. Graphite, is a crystal allotrope of carbon, exhibiting good electric conductivity due to the delocalized $\pi$-electrons between carbon hexagonal lattice layers; Magnetite, a natural mineral of iron ores, exhibits electrical conductivity attributed to electron exchange between iron ions in different oxidation states within the crystal lattice.

We first characterize the conductivity of the fabricated samples (see \textcolor{blue}{Section \ref{Conductivity}} for details), revealing a direct positive correlation between the addition of conductive materials and enhanced conductivity. The results are shown in \textcolor{blue}{Figure~\ref{fig:conductivity_implementation}a}. We note that graphite's inclusion resulted in a more pronounced increase in conductivity compared to magnetite, reflective of its higher electrical conductivity (graphite: $3.3\times10^2$ to $2\times10^5$ S/m, magnetite: 0.1 S/m). We observe a stark rise in conductivity between graphite concentrations of 5\% to 8\%, marking the percolation threshold--- a point at which our interactive material transitions from a nearly insulator to a conductor due to the formation of a continuous path for electron flow. The resistivity and conductivity of pure chlorella are characterized at $1.86\times10^8 \Omega \text{m}$ and $5.37\times10^{-9}$ S/m, respectively.

Complementary to conductivity characterization, we assessed the mechanical integrity of our interactive materials through flexural strength analysis. We conducted a three-point bending test across varying ratios of graphite and magnetite (see \textcolor{blue}{Section \ref{Flexural}} for details). We observe a marginal decline in flexural strength as the proportion of conductive fillers increased, as shown in \textcolor{blue}{Figure~\ref{fig:conductivity_implementation}b}.

For the decoding of conductivity, we implement single-point bio-impedance sensing \cite{waghmare_z-ring_2023} using a vector network analyzer (VNA), the sensing setup is shown in \textcolor{blue}{Figure A\ref{fig:bioimpedance_setup}}. A VNA can transmit a signal into an antenna, and subsequently analyze the reflected signal to measure the antenna's impedance. We apply this principle and use the VNA (LiteVNA) to channel low-power (limit to 5 dBmW, safe for humans) radio frequency (RF) waves through a contact electrode into the user's finger. The impedance mismatch between the port and the finger causes a portion of the signal to reflect back. Once the electrode establishes contact with the skin, it becomes defined and relatively stable. Interaction with external conductive materials by the user introduces additional impedance variations as the signal extends flow from the sensor to these materials. By quantifying changes in the reflected signal, i.e., S11, we can identify the conductive interactive materials touched by the user.

During our real-world measurements, we positioned the electrode on the proximal phalange of the user's contact finger to maintain consistency with our simulation protocols. The measurement is conducted under natural interactions, the interactive materials are placed across different substrates, such as wood and plastic tabletops, and the user can touch them at various pitch angles to simulate real-world usage scenarios of the interactive materials under diverse conditions, thereby stress-testing the robustness. The impedance sensor is configured to transmit signal sweeps ranging from 1 MHz to 500 MHz, executing 30 sweeps per second and reading 51 data points with each sweep. 

The S11 parameter readings collected cross-session from a single user (each session consisted of 5 repetitions of each conductive state, and each repetition consisted of 1 measurement of 2-second duration), follow the same trend as our theoretical model and EM simulation results, as shown in \textcolor{blue}{Figure~\ref{fig:conductivity_implementation}c}. Each conductive state shows a distinct impedance demarcation, offering good SNR.

The culmination of these efforts, the coherence presented from theoretical underpinnings to empirical validations, underscores the practicability of leveraging electrical properties as a medium for information encoding within materials and deploying wearable impedance sensors for decoding.

\section{Magnetic Encoding and Sensing}

In this section, we introduce another new information dimension for designing interactive materials using magnetism. Given that human skin is devoid of inherent magnetoreceptors, we utilize commercial magnetometers for sensing which are available in ubiquitous mobile and wearable devices. We again begin by developing a theoretical model, followed by simulations and experimental validation.

\subsection{Theory}
We develop a model for measuring magnetization with a wearable magnetometer as the user approaches or makes contact with the interactive materials. We develop closed form equations using classical theories of material magnetism, saturation magnetization, and magnetic field strength.

Magnetism manifests in several basic types in classic theory, distinguished by magnetic susceptibility, i.e., its response to external magnetic fields. However, only a few of these types can be attracted to magnetic fields: paramagnetism (e.g., aluminum, lithium), ferromagnetism (e.g., iron, nickel), and ferrimagnetism (e.g., magnetite). While scientists have discovered new types of magnetism in recent years, we will not discuss those here. We narrow our focus to Ferromagnetism and Ferrimagnetism owing to their ability to sustain magnetization in the absence of an external magnetic field, which is necessary to encode persistent data in our IMs.

To develop our  model, we first derive the saturation magnetization $M_s$, to indicate the maximum magnetization a material can achieve under an applied magnetic field. The density of the material $D$ relates to the number of magnetic atoms per unit volume $N$ as $N = \frac{DN_A}{M}$, where $N_A$ is Avogadro's number $6.02\times10^{23} \text{mol}^{-1}$, and $M$ is the molar mass of the material. Saturation magnetization $M_{sat}$ is fundamentally linked to the number of Bohr magnetons per atom $n_B$, density $D$, and molar mass $M$ of the material:

\begin{equation}
    M_{sat} = n_B \mu_B N = \frac{n_B \mu_B D (6.02\times10^{23} \text{mol}^{-1})}{M}
\label{eq:saturation-magnetization}
\end{equation}
where $\mu_B$ is the magnetic moment of its constituents (Bohr magnetons) equals $9.27\times10^{-24} \text{A-m}^2$ in this case.

Using an example from the previous section, assume we have a material with dimensions of $20\times20\times5 \text{mm}$ with 80\% chlorella and 20\% Fe3O4. We know the density of chlorella and Fe3O4--- $0.41\times10^6 \text{g/m}^3$ and $5.17\times10^6 \text{g/m}^3$, respectively. It is noteworthy that chlorella exhibits diamagnetism rendering its magnetic moment negligible in our context. Fe3O4 encompasses both Fe2+ and Fe3+ ions, composing its full formula as $Fe^{2+}(Fe^{3+})_2(O^{2-})_4$. Within this structure, Fe2+ ions occupy sites within the octahedral lattice, while Fe3+ can reside in both octahedral and tetrahedral sites. We assume an equal distribution of Fe3+ ions across tetrahedral and octahedral lattices, so their net magnetic spin is zero. A Fe2+ cation has 4 to 6.7 Bohr Magnetons \cite{parks_magnetic_1968}, thus we have
\begin{align}
    M_{sat} &= \frac{4 (9.27\times10^{-24} \text{A-m}^2)(1.03\times10^6 \text{g/m}^3)(6.02\times10^{23} \text{mol}^{-1})}{231.533 \text{ g/mol}} \nonumber \\
    &= 0.997\times10^{5} \text{A/m}
\label{eq:saturation-magnetization_calculation}
\end{align}

We know the saturation magnetic induction $B_{sat} = \mu_0 M_{sat}$ where $\mu_0=4\pi\times10^{-7}\text{N/A}^2$, now we arrive at our final value:
\begin{equation}
    B_{sat} = (4\pi\times10^{-7}\text{N/A}^2)(0.997\times10^{5} \text{A/m})=0.125 \text{ T}
\end{equation}


Further, we quantify the magnetic flux density $B$ at a distance $X$ from the magnetic interactive material surface, crucial for understanding the noise margin of decoding the magnetization through a magnetometer. We again assume the magnetic interactive material is a solid cuboid with length $L$, width $W$, and thickness $t$. The magnetic flux density is governed by the geometry of the interactive material and its residual magnetization $B_r$, as described by the following equation:
\begin{align}
    B(X) = \frac{B_r}{\pi} \biggr[&\arctan \left(\frac{LW}{2X \sqrt{4X^2+L^2+W^2}}\right) \nonumber \\
    &- \arctan\left(\frac{LW}{2(t+X) \sqrt{4(t+X)^2+L^2+W^2}}\right) \biggr]
\label{eq:magnetic-induction-rectangular}
\end{align}

This equation enables the calculation and simulation of the magnetic field strength when users make contact with the interactive materials at varying distances from the material surface. 




\begin{figure*}[t]
    \centering
    \includegraphics[width=\textwidth]{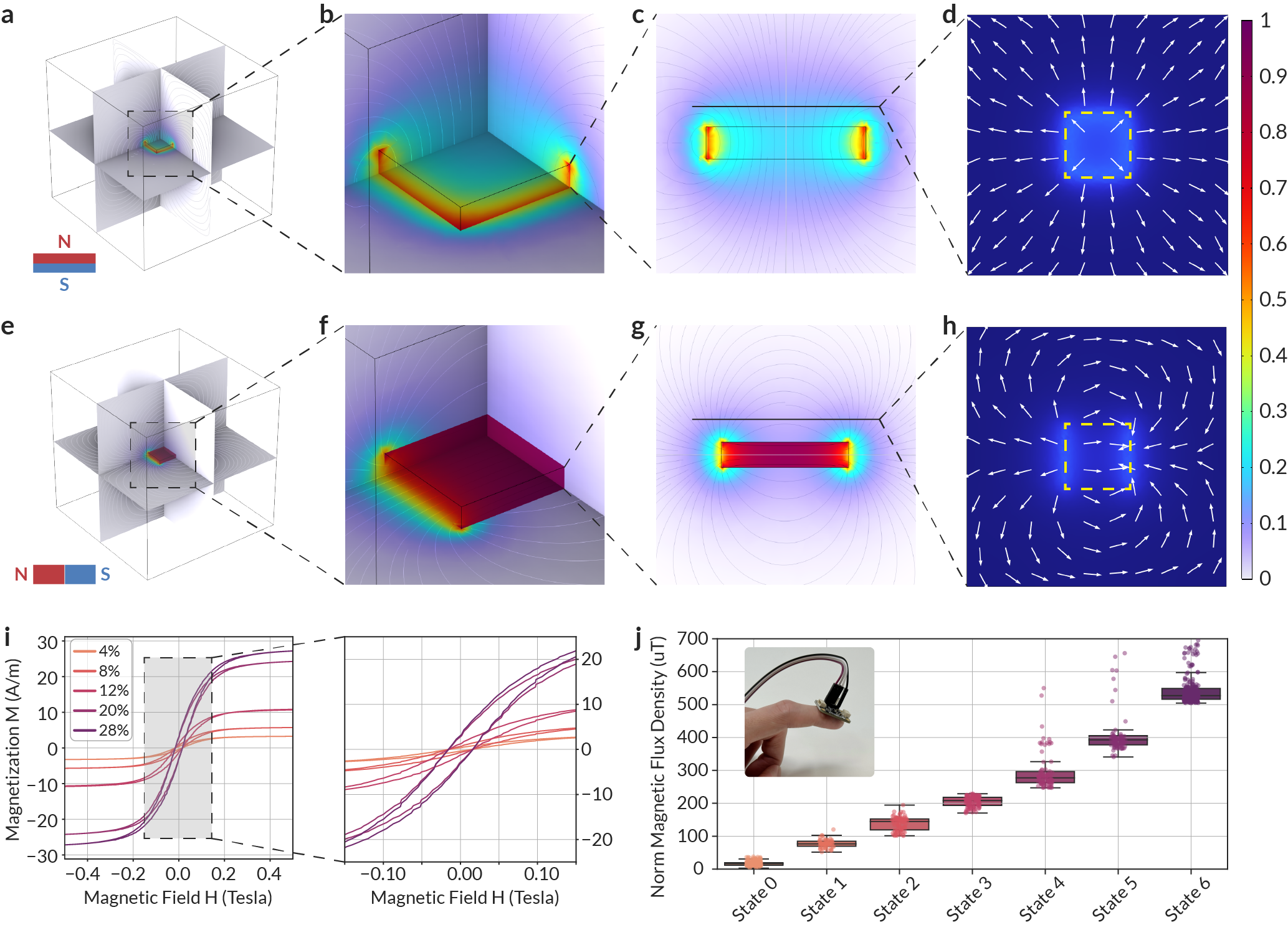}
    \caption{Magnetism. a-h, Simulated magnetic fields of IMs with the poles along the top/bottom bases (a-d) and on opposing faces (e-h), with zoom-in (b, f), front (c, g), and top (d, h) view showing the magnetic field around the magnetic IMs. The simulation results show that top/bottom-oriented poles create a more uniformly distributed magnetic field around the material. i, Characterized magnetic hysteresis loops of magnetic IMs with different proportions of magnetite Fe3O4, showing increasing magnetite content enhances all magnetic properties. j, Measurements from a wearable magnetometer when user's finger is in contact with external IMs of corresponding magnetic states (data is presented as mean (SD) of $N=3\times10\times1\times10=300$ per state).}
    \label{fig:magnetism}
\end{figure*}

\subsection{Simulation}
To use the material's magnetism for information encoding, we need to understand how to program the orientations of magnetic poles effectively. To this end, we conducted finite element analysis (FEA) using COMSOL Multiphysics to optimize the arrangement of the North and South poles—either along the top and bottom bases or on opposing faces. We also aim to assess the remnant magnetism at various proximities to the magnetic IM in the simulation.

In our simulation setup, we modeled a rectangular cuboid with dimensions of $20\times20\times5 \text{mm}$ in COMSOL, representing the magnetic IM. We assigned magnetic poles on either its top/bottom bases and on two opposite side faces, as shown in \textcolor{blue}{Figure~\ref{fig:magnetism}a, e}. The cuboid is placed at the spatial center of a significantly larger air region with dimensions of $100\times100\times100 \text{mm}$ to capture the magnetic field dispersion (\textcolor{blue}{Figure A\ref{fig:comsol_setup}}). The relative permeability of air is set to 1. The recoil permeability and remanent flux density are both normalized to 1 for simplicity of comparison. The mesh of the interactive material is set to 0.4 mm with a curvature factor of 0.4 and a maximum element growth rate of 1.4. We customize the mesh to be less detailed at the air region to reduce memory usage, the outermost air region has an element size of 5mm. The simulation took approximately 1 minute on a 14-core CPU and a single RTX 4070Ti GPU.

The results of each pole orientation are shown in \textcolor{blue}{Figure~\ref{fig:magnetism}b-d, f-h}. Side-facing poles (\textcolor{blue}{Figure~\ref{fig:magnetism}e-h}) mainly confined the magnetic field within the material and vicinity of the poles. Conversely, top/bottom-oriented poles (\textcolor{blue}{Figure~\ref{fig:magnetism}a-d}) resulted in a more uniformly distributed magnetic field around the material, especially in the region above and under the top/bottom bases. This is more aligned with our objective of achieving a magnetic field for magnetometer decoding with minimal variance or noise margin.

We then examine the magnetic flux density across the lines that are perpendicular to the top face of the interactive material. As shown in \textcolor{blue}{Figure A\ref{fig:magnetism_comsol_z}}, the magnetic flux density varies 81.52\% between the distance of 2.5 mm to 10 mm above the top face. We consider this variance in contact distance as the noise region for magnetization decoding. 

We assume the boundaries of $B_r$ are from $5\times10^{-5} \text{ T}$ (Earth's magnetic field at the surface) to $1\times10^{-2} \text{ T}$ (the magnetic field strength of a fridge magnet) as a reasonable range of magnetization in our daily life. Under these assumptions, by tuning the magnetism of the IM while keeping the dimensions identical, we show that it is possible to encode approximately 8 states, equivalent to 3 bits, as expressed by the following equation:

\begin{equation}
    \log_{1.81}{(1\times10^{-2} \text{ T} \div 5\times10^{-5} \text{ T})} \approx 8 \text{ states} \Leftrightarrow \log_{2}{8} = 3 \text{ bits}
    \label{eq:magnetism_result}
\end{equation}

\subsection{Implementation}
We fabricated IMs using a blend of chlorella and varying proportions (0\%-28\%) of magnetite Fe3O4. We characterized the magnetic hysteresis curves (see \textcolor{blue}{Section \ref{Magnetization}} for details) of our magnetic IMs to evaluate saturation magnetization, retentivity, coercivity, and permeability across different Fe3O4 ratios (\textcolor{blue}{Figure~\ref{fig:magnetism}i}). As expected, increasing the magnetite content enhanced all the magnetic properties. We note that the level of magnetic saturation and retentivity of our magnetic IMs are not to the extent of prior studies \cite{iyer_3d_2017, nisser_mixels_2022}, as they used synthetic ferromagnetic materials while we used natural ferrimagnetism magnetite.

The magnetic IMs are then magnetized using either a strong magnet \cite{iyer_3d_2017} or an electromagnet \cite{nisser_mixels_2022} to reach their saturation magnetization. This ensured the magnetic IMs retain their magnetization at a specific retentive level once the external magnetic source is removed, for stable information encoding.

Following our calculation in Eq.\ref{eq:magnetism_result}, the magnetic field strength is segmented into intervals every 1.81 folds (i.e., the read margin of each state is 1.81 times larger than the previous one) to ensure a good SNR for each magnetization state. It is noteworthy that a magnetic IM with 28\% Fe3O4 could not reach the theoretical upper limit of magnetization encoding post-external field removal. To counteract this, we borrowed a commercial magnetic sheet to simulate \textit{State 6} of the magnetic IM states for demonstration purposes.

We decode the embedded magnetic data using a 3-axis magnetometer (Memsic MMC5603) integrated within most modern wearable devices. We placed the magnetometer on the distal phalange of the interacting finger to capture the increase in magnetic field strength as the user's finger approaches the magnetic IM surface. We configured the magnetometer to a 10 Hz sampling rate. During the measurement, the user touches the magnetic IM using the defined finger at various gestures and pitch angles to maximize the variance in contact distance. However, since the sensor was placed on the distal phalange, the contact distance variations remained within acceptable ranges for accurate decoding. 

We conducted three sessions of measurement from one participant; each session consisted of 10 repetitions of each magnetic IM with pre-coded magnetic states, and each repetition consisted of one measurement of 1-second duration. The results are shown in \textcolor{blue}{Figure~\ref{fig:magnetism}j}, each magnetism state shows a distinct flux density demarcation even after taking into account the variance in contact distance. The experiments show we can robustly encode 3 bits of information using magnetism.
\section{Surface Encoding and Sensing}

\begin{figure*}[t]
    \centering
    \includegraphics[width=\textwidth]{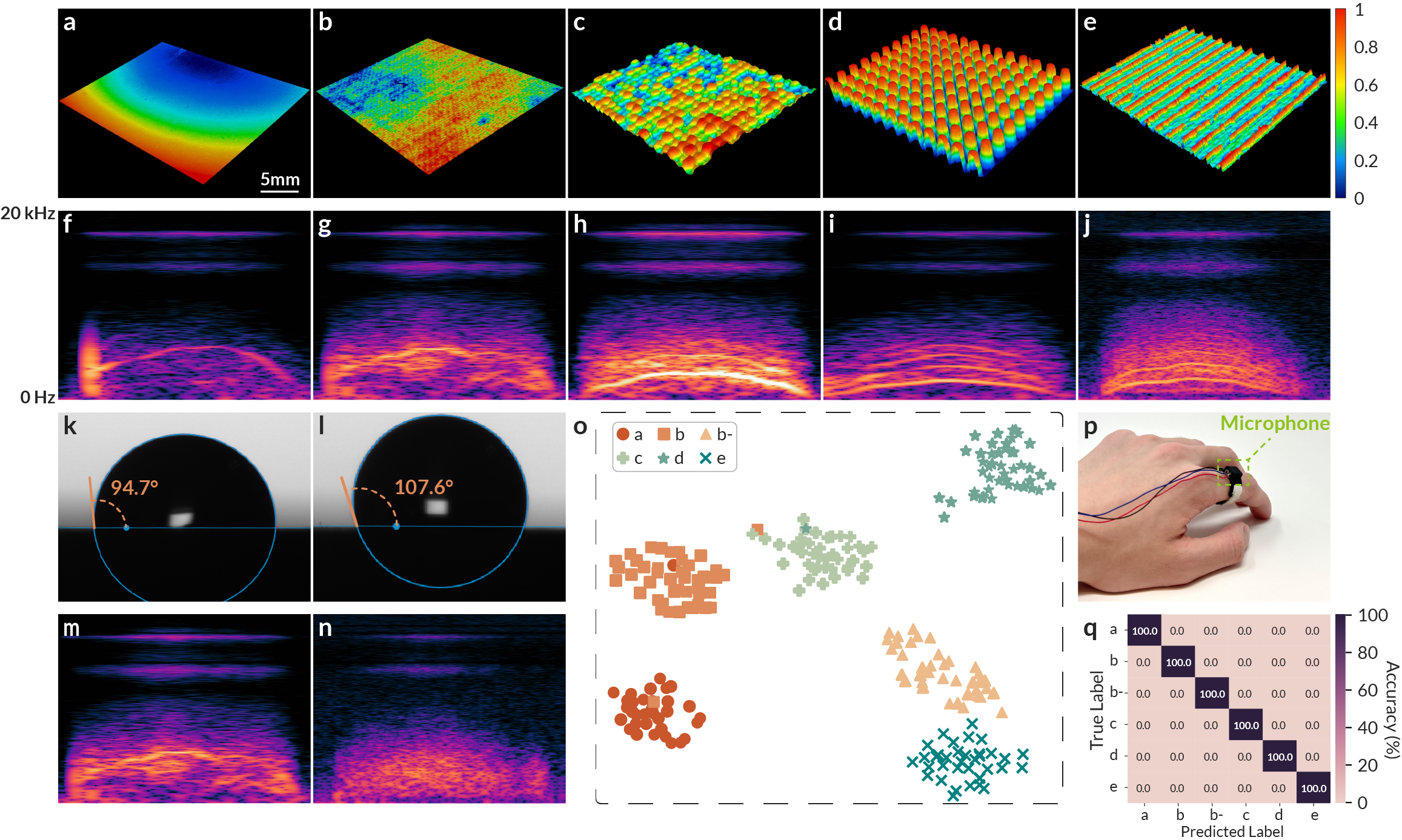}
    \caption{Surface properties. a-j, Micro surface feature images of a set of 5 texture primitives between 200 $\mu$m and 1mm (a-e), and corresponding sound spectrogram measured from a wearable microphone when swiping the IMs (f-j): smooth (a, f), velvet (b, g), rough (c, h), grippy (d, i), and subtle guide (d, j). k-n, Characterized contact angles of IMs with `velvet' surface texture with different surface energy levels, and corresponding sound spectrogram. o, t-SNE plot from our sound measurements recorded by a microphone at the proximal phalange, showing the separation of clusters corresponding to each surface texture and energy. p, Wearable microphone setup. q, Confusion matrix for classifying swipes on different IM surfaces, achieving 100\% accuracy when the window size covers a full swipe. Note that the y-axis is on a mel scale for all sound spectrograms.}
    \label{fig:surface}
\end{figure*}


In addition to encoding information in electromagnetic properties, we explore encoding information in textures inspired by animals like spiders that can detect minute surface vibrations. While human skin does not possess the high-sensitivity mechanoreceptors found in spiders or rodent whiskers, the prevalence of microphones in contemporary wearable devices offers a promising way of capturing the subtle vibrations traveling through air or human skin caused by human touch or swipe on material surfaces. By translating these acoustic vibrations into a form of embedded information, we can leverage textures to enable interactions.

\subsection{Surface Textures}
We selected a spectrum of textural primitives (smooth, velvet, rough, grippy, subtle guide) within the mesoscale (200 $\mu$m to 1mm) for our proof-of-concept leveraging sample designs from Surface I/O, a recent study on functionalizing surface texture for interactions \cite{ding_surface_2023}. At this scale, textures not only provide salient tactile feedback discernible by human fingers but also maintain a subtlety that does not impede finger movement.

The fabrication of these textures employed a variety of techniques including simple heat pressing (a, smooth), sanding (b, velvet) or laser cutting (e, subtle guide) following heat pressing, and 3D printing (c, rough; d, grippy) to achieve the desired surface features. The micro surface feature images of the fabricated textures are captured using a 3D optical profilometer (VR-6000, Keyence, USA) as shown in \textcolor{blue}{Figure~\ref{fig:surface}a-e}, illustrating the differences among them.

In pursuit of enabling ubiquitous interactive materials, we refrain from integrating any electronic components or circuits directly into the material itself. Our objective is to conduct all sensing activities with a user worn device. Instead of employing piezo sensors positioned behind the material to detect vibrations in the material induced by human interaction, as seen in prior research \cite{ding_surface_2023}, we opt for a different approach. We place a 1-axis accelerometer (Knowles BU-21771-000) at the proximal phalange of the interacting finger, repurposing it as a contact microphone to capture vibrations transmitted through the human skin resulting from touch or swipe gestures.

Acoustic signals ranging from 0 to 20 kHz are read from the microphone. We apply post-processing notch filters with a quality factor of 10 at frequencies of 60, 8000, 14000, and 16000 Hz to filter out electric noise.

Our demonstration reveals that a wearable contact microphone at the proximal phalange of the contact finger can capture distinct vibrations produced by human swipes on different surface textures. \textcolor{blue}{Figure~\ref{fig:surface}f-j and Supplementary Video 1} show the visibly different spectrograms of the recorded sound during swipe gestures on each surface.

\subsection{Surface Energy}
Beyond surface textures, surface energy is also significant as it influences how easily other substances can adhere to a surface, opening up a new dimension for encoding information through material surface properties.

We manipulate the surface energy of our textural primitives using a PFC-free waterproofing spray (Grangers, UK), chosen for its efficacy in increasing surface tension of water. The spray works by forming a hydrophobic layer on the surface, effectively increasing its resistance to water and, consequently, changing its surface energy. We chose this sustainable, bluesign-certified surface treatment due to ease of availability for testing, and note that alternative bio-based waterproofing options such as waxes and chitosan-based coatings~\cite{wang_all-biomass-based_2022} could be substituted as well.

For a consistent comparison, we choose the `velvet' texture as our control group in this phase of the experiment. This decision was made considering the fine surface features of other surface textures (e.g., `grippy' and `subtle guide') may undergo some alterations with the application of a hydrophobic layer. 

We first clean the surface and remove contaminants of the `velvet' interactive material using isopropanol (IPA) and deionized (DI) water, then spray the waterproofing solution directly onto the damp `velvet' surface. We remove any surplus liquid with low-lint wipers (Kimwipes, USA) to prevent the formation of an overly thick hydrophobic layer that could obscure the texture's innate features. The treated surfaces are left to dry in a well-ventilated environment for 20 minutes. The modification in surface energy is quantitatively assessed by measuring the contact angle of DI water droplets (see \textcolor{blue}{Section \ref{ContactAngle}} for details), as shown in \textcolor{blue}{Figure~\ref{fig:surface}k, l}.

Employing the same microphone setup described in the preceding section, we observed a distinct acoustic differentiation between surfaces with unmodified (high surface energy) and modified surface (low surface energy) energy levels, as shown in \textcolor{blue}{Figure~\ref{fig:surface}m, n}, respectively.

\subsection{Decoding Surface}
The complexity of theoretically deriving or modeling the friction sounds and sound radiation to the surrounding media due to human swipe necessitates an empirical approach to validate our wearable microphone's sensitivity. We employed t-distributed stochastic neighbor embedding (t-SNE, perplexity=10, default learning rate=200) to project the high-dimensional time series vibration profiles to two-dimensional space, observing distinct clusters of various surface textures and energy states without tuning the hyperparameters, as shown in \textcolor{blue}{Figure~\ref{fig:surface}o}.

We further analyze through a support vector machines classifier with all hyperparameters remaining at default during training, revealing an impressive 100\% accuracy when utilizing complete swipe sound sequences as input embeddings (\textcolor{blue}{Figure~\ref{fig:surface}p}). To enhance the inference speed for the practicality of our system in real-time applications, we also evaluated the classifier's performance on shorter, 0.2-second sound segments, the performance remains high --- $\sim$94\%. These promising results indicate that different textures and energy levels result in distinct resonant frequencies when humans swipe across the surface.

Although developing an accurate physics model for these vibration profiles is challenging, our experiments affirm the feasibility of using surface properties to encode information, at least 3 bits. This approach, combined with the decoding capabilities of ubiquitous microphones in wearable devices, enables an intuitive and natural way for individuals to interact with their surroundings through wearable computing technologies.
\begin{figure*}[t]
    \centering
    \includegraphics[width=\textwidth]{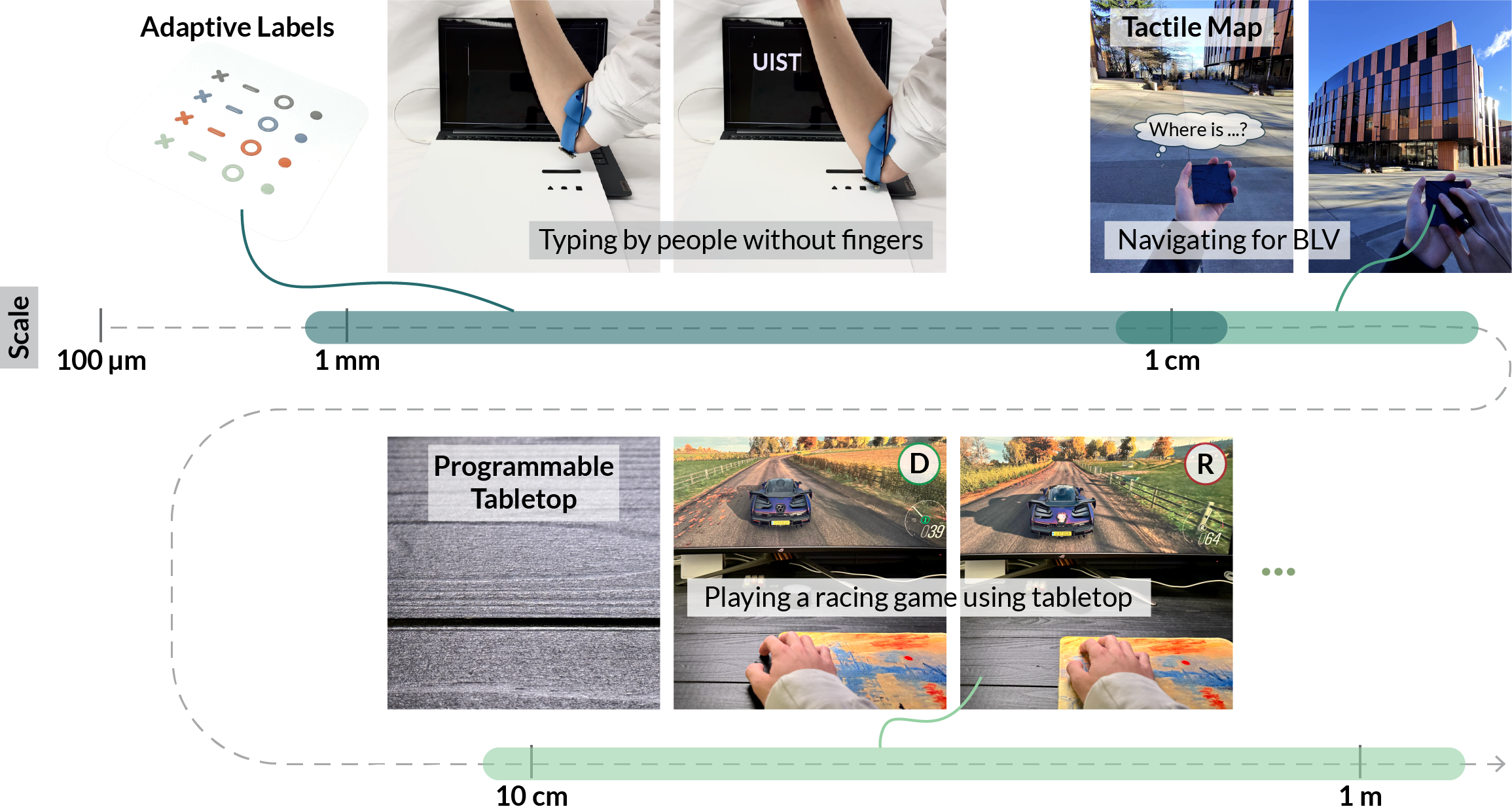}
    \caption{Application overview. A sampling of representative IM application scenarios across geometric scales from sub-milli to meters. The passive, chip-less, tactile interfaces can enable sustainable, ubiquitous interactive systems. We demonstrate a set of customizable label inputs for typing without the need for finger dexterity, tactile maps in which touch triggers audio feedback for BLVs, and programmable surfaces (tabletop and mouse pad) for interactive games, for milli, centi, and decimeter scales, respectively. Note that the programmable tabletop is a Wizard of Oz.}
    \label{fig:application}
\end{figure*}

\section{Envisioned Applications}

We use the techniques developed above to create end-to-end prototypes of interactive materials across a broad range of applications. Each of the application demonstrations described below consists of two components: 1) ubiquitous, passive, chip-less interactive materials with information embedded using conductivity, magnetism, textures or a combination of these; 2) wearable device prototypes equipped with common sensors, including but not limited to magnetometers and microphones, to decode the embedded data. Our systems use these bit patterns as mappings to unique buttons or touch point locations, which can in turn be used to generate audio outputs, update a display, or do a wide variety of other user interaction tasks.

Our representative applications span a spectrum of geometric scales, from millimeters to meters. These applications highlight the practicality of weaving ubiquitous surface interactions directly into the material of everyday objects. We envision a future wherein every object, endowed with information through its inherent material properties, becomes a node in a network of chip-less, power-free, and sustainable interactive surfaces.

\subsection{Millimeter-Scale}
In the realm of everyday interactions, stickers and labels serve as ubiquitous tools for conveying visual information. Beyond their visual utility, specific labels also provide tactile feedback, as exemplified by the Microsoft Surface Adaptive Kit\footnote{\noindent\href{https://www.microsoft.com/en-us/d/surface-adaptive-kit/8rdmhd3kfc3r?activetab=pivot:overviewtab}{https://www.microsoft.com/en-us/d/surface-adaptive-kit/}}, offering essential cues for individuals who are BLV or those without hands or forearms to navigate surfaces, locate keys on a computer, and more.

Inspired by this, we fabricated a set of interactive tactile labels crafted through heat press and laser cutting techniques described above with information encoding in magnetism. By programming the labels to exhibit different states of magnetic flux density and utilizing a magnetometer and an MCU positioned on a user's olecranon and cubital fossa, respectively (see \textcolor{blue}{Figure A\ref{fig:magnetometer_MCU_setup}} for the magnetometer wearable setup), simulating the use conditions for individuals without fingers. The sensor reading was processed on the MCU through simple thresholding, and subsequently mapped to corresponding keyboard keys. The wearable is connected to a laptop to trigger the input. We demonstrate (in \textcolor{blue}{Figure~\ref{fig:application} and Supplementary Video 2}) how such a system can potentially enable the typing of words like "UIST" without the need for finger dexterity.


\subsection{Centimeter-Scale}
Tactile diagrams and maps are regularly used to help convey spatial and conceptual information to BLV individuals. Users report that despite the professionally crafted tactile diagrams, balancing information density with comprehensibility remains challenging \cite{phutane_tactile_2022}. 
Interactive tactile interfaces that integrate audio or haptic feedback triggered by interactions, aim to alleviate the amount of physical clutter intrinsic to traditional tactile diagrams. Existing solutions often rely on placing touchscreens under a regular tactile diagram \cite{melfi_understanding_2020, li_editing_2019, davis_tangiblecircuits_2020} or embedding capacitive touch sensing within already-existing 3D objects but rely on physical wiring between each capacitive sensing element and a microcontroller \cite{holloway_accessible_2018, cheng_swellsense_2023}. However, these approaches to creating interactive tactile interfaces have a high overhead in setting up the system and lack customization options for reconfigurability.

In contrast, as shown in \textcolor{blue}{Figure~\ref{fig:application} and Supplementary Video 3}, we fabricated an interactive tactile map through 3D printing, with each tactile element (i.e., buildings and roads) encoded with information via electrical conductivity. This interactive map itself is completely chip-less and power-free. We show that a `BLV' user equipped with a bio-impedance sensor (see \textcolor{blue}{Figure A\ref{fig:bioimpedance_setup}} for the bio-impedance wearable setup) can touch different tactile elements on the map to receive feedback (audio in this case) from the wearables, which facilitates navigation for BLV (the voice navigation in Supplementary Video 3 was re-recorded separately for better quality). The bio-impedance sensor was connected to a laptop with a custom Python script running for data processing and decoding; decoding was accomplished by thresholding from 80 MHz to 180 MHz and a majority voting. It is worth noting that although the pre-embedded bits within each tactile element are fixed, the corresponding information or feedback it triggers could be easily reconfigured and is fully customizable. For example, mapping `10101' from "The CS building is on the xxx St, on the side closer to the water" to "UIST 2024 is in Pittsburgh", allowing for dynamic updates and personalization to suit various informational needs.

\subsection{Decimeter-Scale and beyond}

Mark Weiser's vision of ubiquitous computing proposed the integration of everyday computational objects across three scales: tabs (inch-scale), pads (foot-scale), and boards (yard-scale). Echoing this vision, we extend the application of interactive materials to the pad scale, envisioning electronic pads that seamlessly integrate into our workspaces, akin to how we spread papers on a desk.

Beyond the millimeter and centimeter-scale demonstrations, we also envision the potential for transforming everyday surfaces like entire tabletops and mouse pads into interactive interfaces. By encoding information through surface textures, we capture the distinct vibration profiles generated by user interactions. For instance, in \textcolor{blue}{Figure~\ref{fig:application}}, shows how a user could control a racing game by swiping across a `rough' textured tabletop to accelerate or a `velvet' textured mouse pad to brake. This deci-scale application exemplifies the broader implications of our work, indicating a potential future where the mundane surfaces around us become conduits for rich, interactive experiences.
\section{Discussion \& Conclusion}


In this paper, we introduce novel methods to create backyard-biodegradable interactive materials with information embedded into their material properties. We draw inspiration from nature for three distinct information encoding strategies. We then use a combination of physics models, computational simulations, and real-world implementations to characterize the performance of each. We then demonstrate how off-the-shelf sensors integrated into wearables can extend our human senses to perceive this data and enable applications like customizable buttons or tactile maps. Furthermore, we demonstrate our prototypes degrade naturally outdoors and we show through a comparative LCA that our approach would have significant environmental benefits in a future where interactive surfaces are ubiquitous.

While our work paves a new way for creating sustainable interactive interfaces, it also reveals multiple opportunities for future research. First, a critical aspect of broadening the practicability of these biodegradable interactive materials involves their aesthetic appearances, such as color, with everyday objects. However, it is challenging to achieve a diverse color palette with black graphite or magnetite. We found that by mixing any color of the base material with black graphite or magnetite, resulting in a black hue, the resulting IM looks almost identically black. Improvements could be made by exploring alternative colors or material blends using materials such as bornite (iridescence) or malachite (green). Furthermore, this work currently demonstrates the use of three natural material properties for information encoding, future work may reveal a broader range of material properties used for information encoding, such as incorporating thermal conductivity and hardness. Investigating these additional properties could exponentially increase the capacity for information encoding within materials.

The second promising direction for future work is the development of flexible IMs that can conform to the non-flat surfaces commonly encountered in daily life. By experimenting with base materials such as flax or bacterial cellulose, we can further apply interactive interfaces to a broader array of objects and settings.

Third, we show the independent use of off-the-shelf sensors for decoding different material properties. A potential future work lies in integrating these sensors into a singular, functional, wireless wearable device. Such a device would facilitate conducting further user studies in natural, everyday environments, as it would eliminate the constraints imposed by wired connections and power requirements on wearables, thereby exploring the vast design space of interactive materials.

Fourth, in this work we strategically positioned all sensors on the dorsal aspect of the finger to avoid interference with the palmar side. The existing off-the-shelf sensors are bulky and enclosed in rigid packages, necessitating this placement to preserve natural tactile feedback and manual dexterity during daily activities. Consequently, sensor readings are influenced by variables such as contact angle, force, and distance, which we consider as noise margins in the context of information encoding. However, the rapid advancement in electronic skin (e-skin) offers promising alternatives that could revolutionize our approach in the future. These e-skins \cite{tee_skin-inspired_2015, wang_neuromorphic_2023} are capable of replicating and even enhancing the nuanced sensory feedback and receptions of natural skin by incorporating soft sensors and transistors within a thin, stretchable film. The innovation in e-skin promises to increase the information density that can be encoded within IMs, as they enable sensors to direct contact with interactive surfaces via e-skins applied to the finger pulp, while still maintaining hand dexterity.

In summary, this work introduces a novel paradigm of creating surface interaction with multiple avenues for further exploration. We hope this research will not only empower designers within the HCI and Ubicomp communities to create new application scenarios-from daily interactions to accessibility- but also stimulate further investigation into passive, chip-less interactive materials, toward the vision of the Internet of Materials.


\begin{acks}
\end{acks}

\bibliographystyle{ACM-Reference-Format}
\bibliography{references}


\begin{thebibliography}{60}


\ifx \showCODEN    \undefined \def \showCODEN     #1{\unskip}     \fi
\ifx \showDOI      \undefined \def \showDOI       #1{#1}\fi
\ifx \showISBNx    \undefined \def \showISBNx     #1{\unskip}     \fi
\ifx \showISBNxiii \undefined \def \showISBNxiii  #1{\unskip}     \fi
\ifx \showISSN     \undefined \def \showISSN      #1{\unskip}     \fi
\ifx \showLCCN     \undefined \def \showLCCN      #1{\unskip}     \fi
\ifx \shownote     \undefined \def \shownote      #1{#1}          \fi
\ifx \showarticletitle \undefined \def \showarticletitle #1{#1}   \fi
\ifx \showURL      \undefined \def \showURL       {\relax}        \fi
\providecommand\bibfield[2]{#2}
\providecommand\bibinfo[2]{#2}
\providecommand\natexlab[1]{#1}
\providecommand\showeprint[2][]{arXiv:#2}

\bibitem[Abowd(2020)]%
        {abowd_internet_2020}
\bibfield{author}{\bibinfo{person}{Gregory~D. Abowd}.} \bibinfo{year}{2020}\natexlab{}.
\newblock \showarticletitle{The {Internet} of {Materials}: {A} {Vision} for {Computational} {Materials}}.
\newblock \bibinfo{journal}{\emph{IEEE Pervasive Computing}} \bibinfo{volume}{19}, \bibinfo{number}{2} (\bibinfo{date}{April} \bibinfo{year}{2020}), \bibinfo{pages}{56--62}.
\newblock
\showISSN{1558-2590}
\urldef\tempurl%
\url{https://doi.org/10.1109/MPRV.2020.2982475}
\showDOI{\tempurl}
\newblock
\shownote{Conference Name: IEEE Pervasive Computing}.


\bibitem[Anderson et~al\mbox{.}(2022)]%
        {anderson_treehouse_2022}
\bibfield{author}{\bibinfo{person}{Thomas Anderson}, \bibinfo{person}{Adam Belay}, \bibinfo{person}{Mosharaf Chowdhury}, \bibinfo{person}{Asaf Cidon}, {and} \bibinfo{person}{Irene Zhang}.} \bibinfo{year}{2022}\natexlab{}.
\newblock \bibinfo{title}{Treehouse: {A} {Case} {For} {Carbon}-{Aware} {Datacenter} {Software}}.
\newblock
\newblock
\urldef\tempurl%
\url{https://doi.org/10.48550/arXiv.2201.02120}
\showDOI{\tempurl}
\newblock
\shownote{arXiv:2201.02120 [cs]}.


\bibitem[Apple({[n.\,d.]})]%
        {apple_accessibility_nodate}
\bibfield{author}{\bibinfo{person}{Apple}.} \bibinfo{year}{[n.\,d.]}\natexlab{}.
\newblock \bibinfo{title}{Accessibility}.
\newblock
\newblock
\urldef\tempurl%
\url{https://developer.apple.com/design/human-interface-guidelines/accessibility}
\showURL{%
\tempurl}


\bibitem[Arora et~al\mbox{.}(2021)]%
        {arora_mars_2021}
\bibfield{author}{\bibinfo{person}{Nivedita Arora}, \bibinfo{person}{Ali Mirzazadeh}, \bibinfo{person}{Injoo Moon}, \bibinfo{person}{Charles Ramey}, \bibinfo{person}{Yuhui Zhao}, \bibinfo{person}{Daniela~C. Rodriguez}, \bibinfo{person}{Gregory~D. Abowd}, {and} \bibinfo{person}{Thad Starner}.} \bibinfo{year}{2021}\natexlab{}.
\newblock \showarticletitle{{MARS}: {Nano}-{Power} {Battery}-free {Wireless} {Interfaces} for {Touch}, {Swipe} and {Speech} {Input}}. In \bibinfo{booktitle}{\emph{The 34th {Annual} {ACM} {Symposium} on {User} {Interface} {Software} and {Technology}}} \emph{(\bibinfo{series}{{UIST} '21})}. \bibinfo{publisher}{Association for Computing Machinery}, \bibinfo{address}{New York, NY, USA}, \bibinfo{pages}{1305--1325}.
\newblock
\showISBNx{978-1-4503-8635-7}
\urldef\tempurl%
\url{https://doi.org/10.1145/3472749.3474823}
\showDOI{\tempurl}


\bibitem[Arora et~al\mbox{.}(2018)]%
        {arora_saturn_2018}
\bibfield{author}{\bibinfo{person}{Nivedita Arora}, \bibinfo{person}{Steven~L. Zhang}, \bibinfo{person}{Fereshteh Shahmiri}, \bibinfo{person}{Diego Osorio}, \bibinfo{person}{Yi-Cheng Wang}, \bibinfo{person}{Mohit Gupta}, \bibinfo{person}{Zhengjun Wang}, \bibinfo{person}{Thad Starner}, \bibinfo{person}{Zhong~Lin Wang}, {and} \bibinfo{person}{Gregory~D. Abowd}.} \bibinfo{year}{2018}\natexlab{}.
\newblock \showarticletitle{{SATURN}: {A} {Thin} and {Flexible} {Self}-powered {Microphone} {Leveraging} {Triboelectric} {Nanogenerator}}.
\newblock \bibinfo{journal}{\emph{Proceedings of the ACM on Interactive, Mobile, Wearable and Ubiquitous Technologies}} \bibinfo{volume}{2}, \bibinfo{number}{2} (\bibinfo{date}{July} \bibinfo{year}{2018}), \bibinfo{pages}{60:1--60:28}.
\newblock
\urldef\tempurl%
\url{https://doi.org/10.1145/3214263}
\showDOI{\tempurl}


\bibitem[Asakawa et~al\mbox{.}(2017)]%
        {asakawa_fingertip_2017}
\bibfield{author}{\bibinfo{person}{Deanna~S. Asakawa}, \bibinfo{person}{George~H. Crocker}, \bibinfo{person}{Adam Schmaltz}, {and} \bibinfo{person}{Devin~L. Jindrich}.} \bibinfo{year}{2017}\natexlab{}.
\newblock \showarticletitle{Fingertip forces and completion time for index finger and thumb touchscreen gestures}.
\newblock \bibinfo{journal}{\emph{Journal of Electromyography and Kinesiology}}  \bibinfo{volume}{34} (\bibinfo{date}{June} \bibinfo{year}{2017}), \bibinfo{pages}{6--13}.
\newblock
\showISSN{1050-6411}
\urldef\tempurl%
\url{https://doi.org/10.1016/j.jelekin.2017.02.007}
\showDOI{\tempurl}


\bibitem[Blakemore(1975)]%
        {blakemore_magnetotactic_1975}
\bibfield{author}{\bibinfo{person}{Richard Blakemore}.} \bibinfo{year}{1975}\natexlab{}.
\newblock \showarticletitle{Magnetotactic {Bacteria}}.
\newblock \bibinfo{journal}{\emph{Science}} \bibinfo{volume}{190}, \bibinfo{number}{4212} (\bibinfo{date}{Oct.} \bibinfo{year}{1975}), \bibinfo{pages}{377--379}.
\newblock
\urldef\tempurl%
\url{https://doi.org/10.1126/science.170679}
\showDOI{\tempurl}
\newblock
\shownote{Publisher: American Association for the Advancement of Science}.


\bibitem[Campbell et~al\mbox{.}(2023)]%
        {campbell_progress_2023}
\bibfield{author}{\bibinfo{person}{Ian~R. Campbell}, \bibinfo{person}{Meng-Yen Lin}, \bibinfo{person}{Hareesh Iyer}, \bibinfo{person}{Mallory Parker}, \bibinfo{person}{Jeremy~L. Fredricks}, \bibinfo{person}{Kuotian Liao}, \bibinfo{person}{Andrew~M. Jimenez}, \bibinfo{person}{Paul Grandgeorge}, {and} \bibinfo{person}{Eleftheria Roumeli}.} \bibinfo{year}{2023}\natexlab{}.
\newblock \showarticletitle{Progress in {Sustainable} {Polymers} from {Biological} {Matter}}.
\newblock \bibinfo{journal}{\emph{Annual Review of Materials Research}} \bibinfo{volume}{53}, \bibinfo{number}{Volume 53, 2023} (\bibinfo{date}{July} \bibinfo{year}{2023}), \bibinfo{pages}{81--104}.
\newblock
\showISSN{1531-7331, 1545-4118}
\urldef\tempurl%
\url{https://doi.org/10.1146/annurev-matsci-080921-083655}
\showDOI{\tempurl}
\newblock
\shownote{Publisher: Annual Reviews}.


\bibitem[Cheng et~al\mbox{.}(2023a)]%
        {cheng_functional_2023}
\bibfield{author}{\bibinfo{person}{Tingyu Cheng}, \bibinfo{person}{Taylor Tabb}, \bibinfo{person}{Jung~Wook Park}, \bibinfo{person}{Eric~M Gallo}, \bibinfo{person}{Aditi Maheshwari}, \bibinfo{person}{Gregory~D. Abowd}, \bibinfo{person}{Hyunjoo Oh}, {and} \bibinfo{person}{Andreea Danielescu}.} \bibinfo{year}{2023}\natexlab{a}.
\newblock \showarticletitle{Functional {Destruction}: {Utilizing} {Sustainable} {Materials}’ {Physical} {Transiency} for {Electronics} {Applications}}. In \bibinfo{booktitle}{\emph{Proceedings of the 2023 {CHI} {Conference} on {Human} {Factors} in {Computing} {Systems}}} \emph{(\bibinfo{series}{{CHI} '23})}. \bibinfo{publisher}{Association for Computing Machinery}, \bibinfo{address}{New York, NY, USA}, \bibinfo{pages}{1--16}.
\newblock
\showISBNx{978-1-4503-9421-5}
\urldef\tempurl%
\url{https://doi.org/10.1145/3544548.3580811}
\showDOI{\tempurl}


\bibitem[Cheng et~al\mbox{.}(2023b)]%
        {cheng_swellsense_2023}
\bibfield{author}{\bibinfo{person}{Tingyu Cheng}, \bibinfo{person}{Zhihan Zhang}, \bibinfo{person}{Bingrui Zong}, \bibinfo{person}{Yuhui Zhao}, \bibinfo{person}{Zekun Chang}, \bibinfo{person}{Yejun Kim}, \bibinfo{person}{Clement Zheng}, \bibinfo{person}{Gregory~D. Abowd}, {and} \bibinfo{person}{HyunJoo Oh}.} \bibinfo{year}{2023}\natexlab{b}.
\newblock \showarticletitle{{SwellSense}: {Creating} 2.{5D} interactions with micro-capsule paper}. In \bibinfo{booktitle}{\emph{Proceedings of the 2023 {CHI} {Conference} on {Human} {Factors} in {Computing} {Systems}}} \emph{(\bibinfo{series}{{CHI} '23})}. \bibinfo{publisher}{Association for Computing Machinery}, \bibinfo{address}{New York, NY, USA}, \bibinfo{pages}{1--13}.
\newblock
\showISBNx{978-1-4503-9421-5}
\urldef\tempurl%
\url{https://doi.org/10.1145/3544548.3581125}
\showDOI{\tempurl}


\bibitem[Clarke et~al\mbox{.}(2017)]%
        {clarke_bee_2017}
\bibfield{author}{\bibinfo{person}{Dominic Clarke}, \bibinfo{person}{Erica Morley}, {and} \bibinfo{person}{Daniel Robert}.} \bibinfo{year}{2017}\natexlab{}.
\newblock \showarticletitle{The bee, the flower, and the electric field: electric ecology and aerial electroreception}.
\newblock \bibinfo{journal}{\emph{Journal of Comparative Physiology A}} \bibinfo{volume}{203}, \bibinfo{number}{9} (\bibinfo{date}{Sept.} \bibinfo{year}{2017}), \bibinfo{pages}{737--748}.
\newblock
\showISSN{1432-1351}
\urldef\tempurl%
\url{https://doi.org/10.1007/s00359-017-1176-6}
\showDOI{\tempurl}


\bibitem[Davis et~al\mbox{.}(2020)]%
        {davis_tangiblecircuits_2020}
\bibfield{author}{\bibinfo{person}{Josh~Urban Davis}, \bibinfo{person}{Te-Yen Wu}, \bibinfo{person}{Bo Shi}, \bibinfo{person}{Hanyi Lu}, \bibinfo{person}{Athina Panotopoulou}, \bibinfo{person}{Emily Whiting}, {and} \bibinfo{person}{Xing-Dong Yang}.} \bibinfo{year}{2020}\natexlab{}.
\newblock \showarticletitle{{TangibleCircuits}: {An} {Interactive} {3D} {Printed} {Circuit} {Education} {Tool} for {People} with {Visual} {Impairments}}. In \bibinfo{booktitle}{\emph{Proceedings of the 2020 {CHI} {Conference} on {Human} {Factors} in {Computing} {Systems}}} \emph{(\bibinfo{series}{{CHI} '20})}. \bibinfo{publisher}{Association for Computing Machinery}, \bibinfo{address}{New York, NY, USA}, \bibinfo{pages}{1--13}.
\newblock
\showISBNx{978-1-4503-6708-0}
\urldef\tempurl%
\url{https://doi.org/10.1145/3313831.3376513}
\showDOI{\tempurl}


\bibitem[de~Winkel et~al\mbox{.}(2020)]%
        {de_winkel_battery-free_2020}
\bibfield{author}{\bibinfo{person}{Jasper de Winkel}, \bibinfo{person}{Vito Kortbeek}, \bibinfo{person}{Josiah Hester}, {and} \bibinfo{person}{Przemysław Pawełczak}.} \bibinfo{year}{2020}\natexlab{}.
\newblock \showarticletitle{Battery-{Free} {Game} {Boy}}.
\newblock \bibinfo{journal}{\emph{Proceedings of the ACM on Interactive, Mobile, Wearable and Ubiquitous Technologies}} \bibinfo{volume}{4}, \bibinfo{number}{3} (\bibinfo{date}{Sept.} \bibinfo{year}{2020}), \bibinfo{pages}{111:1--111:34}.
\newblock
\urldef\tempurl%
\url{https://doi.org/10.1145/3411839}
\showDOI{\tempurl}


\bibitem[Ding et~al\mbox{.}(2023)]%
        {ding_surface_2023}
\bibfield{author}{\bibinfo{person}{Yuran Ding}, \bibinfo{person}{Craig Shultz}, {and} \bibinfo{person}{Chris Harrison}.} \bibinfo{year}{2023}\natexlab{}.
\newblock \showarticletitle{Surface {I}/{O}: {Creating} {Devices} with {Functional} {Surface} {Geometry} for {Haptics} and {User} {Input}}. In \bibinfo{booktitle}{\emph{Proceedings of the 2023 {CHI} {Conference} on {Human} {Factors} in {Computing} {Systems}}} \emph{(\bibinfo{series}{{CHI} '23})}. \bibinfo{publisher}{Association for Computing Machinery}, \bibinfo{address}{New York, NY, USA}, \bibinfo{pages}{1--22}.
\newblock
\showISBNx{978-1-4503-9421-5}
\urldef\tempurl%
\url{https://doi.org/10.1145/3544548.3581037}
\showDOI{\tempurl}


\bibitem[Dzidek et~al\mbox{.}(2017)]%
        {dzidek_contact_2017}
\bibfield{author}{\bibinfo{person}{Brygida~M. Dzidek}, \bibinfo{person}{Michael~J. Adams}, \bibinfo{person}{James~W. Andrews}, \bibinfo{person}{Zhibing Zhang}, {and} \bibinfo{person}{Simon~A. Johnson}.} \bibinfo{year}{2017}\natexlab{}.
\newblock \showarticletitle{Contact mechanics of the human finger pad under compressive loads}.
\newblock \bibinfo{journal}{\emph{Journal of The Royal Society Interface}} \bibinfo{volume}{14}, \bibinfo{number}{127} (\bibinfo{date}{Feb.} \bibinfo{year}{2017}), \bibinfo{pages}{20160935}.
\newblock
\urldef\tempurl%
\url{https://doi.org/10.1098/rsif.2016.0935}
\showDOI{\tempurl}
\newblock
\shownote{Publisher: Royal Society}.


\bibitem[Fredricks et~al\mbox{.}(2021)]%
        {fredricks_spirulina-based_2021}
\bibfield{author}{\bibinfo{person}{Jeremy~L. Fredricks}, \bibinfo{person}{Hareesh Iyer}, \bibinfo{person}{Robin McDonald}, \bibinfo{person}{Jeffrey Hsu}, \bibinfo{person}{Andrew~M. Jimenez}, {and} \bibinfo{person}{Eleftheria Roumeli}.} \bibinfo{year}{2021}\natexlab{}.
\newblock \showarticletitle{Spirulina-based composites for {3D}-printing}.
\newblock \bibinfo{journal}{\emph{Journal of Polymer Science}} \bibinfo{volume}{59}, \bibinfo{number}{22} (\bibinfo{year}{2021}), \bibinfo{pages}{2878--2894}.
\newblock
\showISSN{2642-4169}
\urldef\tempurl%
\url{https://doi.org/10.1002/pol.20210683}
\showDOI{\tempurl}
\newblock
\shownote{\_eprint: https://onlinelibrary.wiley.com/doi/pdf/10.1002/pol.20210683}.


\bibitem[Gupta et~al\mbox{.}(2022)]%
        {gupta_act_2022}
\bibfield{author}{\bibinfo{person}{Udit Gupta}, \bibinfo{person}{Mariam Elgamal}, \bibinfo{person}{Gage Hills}, \bibinfo{person}{Gu-Yeon Wei}, \bibinfo{person}{Hsien-Hsin~S. Lee}, \bibinfo{person}{David Brooks}, {and} \bibinfo{person}{Carole-Jean Wu}.} \bibinfo{year}{2022}\natexlab{}.
\newblock \showarticletitle{{ACT}: designing sustainable computer systems with an architectural carbon modeling tool}. In \bibinfo{booktitle}{\emph{Proceedings of the 49th {Annual} {International} {Symposium} on {Computer} {Architecture}}} \emph{(\bibinfo{series}{{ISCA} '22})}. \bibinfo{publisher}{Association for Computing Machinery}, \bibinfo{address}{New York, NY, USA}, \bibinfo{pages}{784--799}.
\newblock
\showISBNx{978-1-4503-8610-4}
\urldef\tempurl%
\url{https://doi.org/10.1145/3470496.3527408}
\showDOI{\tempurl}


\bibitem[Holloway et~al\mbox{.}(2018)]%
        {holloway_accessible_2018}
\bibfield{author}{\bibinfo{person}{Leona Holloway}, \bibinfo{person}{Kim Marriott}, {and} \bibinfo{person}{Matthew Butler}.} \bibinfo{year}{2018}\natexlab{}.
\newblock \showarticletitle{Accessible {Maps} for the {Blind}: {Comparing} {3D} {Printed} {Models} with {Tactile} {Graphics}}. In \bibinfo{booktitle}{\emph{Proceedings of the 2018 {CHI} {Conference} on {Human} {Factors} in {Computing} {Systems}}} \emph{(\bibinfo{series}{{CHI} '18})}. \bibinfo{publisher}{Association for Computing Machinery}, \bibinfo{address}{New York, NY, USA}, \bibinfo{pages}{1--13}.
\newblock
\showISBNx{978-1-4503-5620-6}
\urldef\tempurl%
\url{https://doi.org/10.1145/3173574.3173772}
\showDOI{\tempurl}


\bibitem[Iyer et~al\mbox{.}(2023)]%
        {iyer_fabricating_2023}
\bibfield{author}{\bibinfo{person}{Hareesh Iyer}, \bibinfo{person}{Paul Grandgeorge}, \bibinfo{person}{Andrew~M. Jimenez}, \bibinfo{person}{Ian~R. Campbell}, \bibinfo{person}{Mallory Parker}, \bibinfo{person}{Michael Holden}, \bibinfo{person}{Mathangi Venkatesh}, \bibinfo{person}{Marissa Nelsen}, \bibinfo{person}{Bichlien Nguyen}, {and} \bibinfo{person}{Eleftheria Roumeli}.} \bibinfo{year}{2023}\natexlab{}.
\newblock \showarticletitle{Fabricating {Strong} and {Stiff} {Bioplastics} from {Whole} {Spirulina} {Cells}}.
\newblock \bibinfo{journal}{\emph{Advanced Functional Materials}} \bibinfo{volume}{33}, \bibinfo{number}{40} (\bibinfo{year}{2023}), \bibinfo{pages}{2302067}.
\newblock
\showISSN{1616-3028}
\urldef\tempurl%
\url{https://doi.org/10.1002/adfm.202302067}
\showDOI{\tempurl}
\newblock
\shownote{\_eprint: https://onlinelibrary.wiley.com/doi/pdf/10.1002/adfm.202302067}.


\bibitem[Iyer et~al\mbox{.}(2017)]%
        {iyer_3d_2017}
\bibfield{author}{\bibinfo{person}{Vikram Iyer}, \bibinfo{person}{Justin Chan}, {and} \bibinfo{person}{Shyamnath Gollakota}.} \bibinfo{year}{2017}\natexlab{}.
\newblock \showarticletitle{{3D} printing wireless connected objects}.
\newblock \bibinfo{journal}{\emph{ACM Transactions on Graphics}} \bibinfo{volume}{36}, \bibinfo{number}{6} (\bibinfo{date}{Nov.} \bibinfo{year}{2017}), \bibinfo{pages}{242:1--242:13}.
\newblock
\showISSN{0730-0301}
\urldef\tempurl%
\url{https://doi.org/10.1145/3130800.3130822}
\showDOI{\tempurl}


\bibitem[Iyer et~al\mbox{.}(2022)]%
        {iyer_wind_2022}
\bibfield{author}{\bibinfo{person}{Vikram Iyer}, \bibinfo{person}{Hans Gaensbauer}, \bibinfo{person}{Thomas~L. Daniel}, {and} \bibinfo{person}{Shyamnath Gollakota}.} \bibinfo{year}{2022}\natexlab{}.
\newblock \showarticletitle{Wind dispersal of battery-free wireless devices}.
\newblock \bibinfo{journal}{\emph{Nature}} \bibinfo{volume}{603}, \bibinfo{number}{7901} (\bibinfo{date}{March} \bibinfo{year}{2022}), \bibinfo{pages}{427--433}.
\newblock
\showISSN{1476-4687}
\urldef\tempurl%
\url{https://doi.org/10.1038/s41586-021-04363-9}
\showDOI{\tempurl}
\newblock
\shownote{Publisher: Nature Publishing Group}.


\bibitem[Iyer et~al\mbox{.}(2020)]%
        {iyer_wireless_2020}
\bibfield{author}{\bibinfo{person}{Vikram Iyer}, \bibinfo{person}{Ali Najafi}, \bibinfo{person}{Johannes James}, \bibinfo{person}{Sawyer Fuller}, {and} \bibinfo{person}{Shyamnath Gollakota}.} \bibinfo{year}{2020}\natexlab{}.
\newblock \showarticletitle{Wireless steerable vision for live insects and insect-scale robots}.
\newblock \bibinfo{journal}{\emph{Science Robotics}} \bibinfo{volume}{5}, \bibinfo{number}{44} (\bibinfo{date}{July} \bibinfo{year}{2020}), \bibinfo{pages}{eabb0839}.
\newblock
\urldef\tempurl%
\url{https://doi.org/10.1126/scirobotics.abb0839}
\showDOI{\tempurl}
\newblock
\shownote{Publisher: American Association for the Advancement of Science}.


\bibitem[Jang and Adib(2019)]%
        {jang_underwater_2019}
\bibfield{author}{\bibinfo{person}{Junsu Jang} {and} \bibinfo{person}{Fadel Adib}.} \bibinfo{year}{2019}\natexlab{}.
\newblock \showarticletitle{Underwater backscatter networking}. In \bibinfo{booktitle}{\emph{Proceedings of the {ACM} {Special} {Interest} {Group} on {Data} {Communication}}} \emph{(\bibinfo{series}{{SIGCOMM} '19})}. \bibinfo{publisher}{Association for Computing Machinery}, \bibinfo{address}{New York, NY, USA}, \bibinfo{pages}{187--199}.
\newblock
\showISBNx{978-1-4503-5956-6}
\urldef\tempurl%
\url{https://doi.org/10.1145/3341302.3342091}
\showDOI{\tempurl}


\bibitem[Johnson et~al\mbox{.}(2023)]%
        {johnson_solar-powered_2023}
\bibfield{author}{\bibinfo{person}{Kyle Johnson}, \bibinfo{person}{Vicente Arroyos}, \bibinfo{person}{Amélie Ferran}, \bibinfo{person}{Raul Villanueva}, \bibinfo{person}{Dennis Yin}, \bibinfo{person}{Tilboon Elberier}, \bibinfo{person}{Alberto Aliseda}, \bibinfo{person}{Sawyer Fuller}, \bibinfo{person}{Vikram Iyer}, {and} \bibinfo{person}{Shyamnath Gollakota}.} \bibinfo{year}{2023}\natexlab{}.
\newblock \showarticletitle{Solar-powered shape-changing origami microfliers}.
\newblock \bibinfo{journal}{\emph{Science Robotics}} \bibinfo{volume}{8}, \bibinfo{number}{82} (\bibinfo{date}{Sept.} \bibinfo{year}{2023}), \bibinfo{pages}{eadg4276}.
\newblock
\urldef\tempurl%
\url{https://doi.org/10.1126/scirobotics.adg4276}
\showDOI{\tempurl}
\newblock
\shownote{Publisher: American Association for the Advancement of Science}.


\bibitem[Jung et~al\mbox{.}(2021)]%
        {jung_wrist-wearable_2021}
\bibfield{author}{\bibinfo{person}{Myoung~Hoon Jung}, \bibinfo{person}{Kak Namkoong}, \bibinfo{person}{Yeolho Lee}, \bibinfo{person}{Young~Jun Koh}, \bibinfo{person}{Kunsun Eom}, \bibinfo{person}{Hyeongseok Jang}, \bibinfo{person}{Wonjong Jung}, \bibinfo{person}{Jungmok Bae}, {and} \bibinfo{person}{Jongae Park}.} \bibinfo{year}{2021}\natexlab{}.
\newblock \showarticletitle{Wrist-wearable bioelectrical impedance analyzer with miniature electrodes for daily obesity management}.
\newblock \bibinfo{journal}{\emph{Scientific Reports}} \bibinfo{volume}{11}, \bibinfo{number}{1} (\bibinfo{date}{Jan.} \bibinfo{year}{2021}), \bibinfo{pages}{1238}.
\newblock
\showISSN{2045-2322}
\urldef\tempurl%
\url{https://doi.org/10.1038/s41598-020-79667-3}
\showDOI{\tempurl}
\newblock
\shownote{Number: 1 Publisher: Nature Publishing Group}.


\bibitem[Kim et~al\mbox{.}(2023)]%
        {kim_remote_2023}
\bibfield{author}{\bibinfo{person}{Yongdeok Kim}, \bibinfo{person}{Yiyuan Yang}, \bibinfo{person}{Xiaotian Zhang}, \bibinfo{person}{Zhengwei Li}, \bibinfo{person}{Abraham Vázquez-Guardado}, \bibinfo{person}{Insu Park}, \bibinfo{person}{Jiaojiao Wang}, \bibinfo{person}{Andrew~I. Efimov}, \bibinfo{person}{Zhi Dou}, \bibinfo{person}{Yue Wang}, \bibinfo{person}{Junehu Park}, \bibinfo{person}{Haiwen Luan}, \bibinfo{person}{Xinchen Ni}, \bibinfo{person}{Yun~Seong Kim}, \bibinfo{person}{Janice Baek}, \bibinfo{person}{Joshua~Jaehyung Park}, \bibinfo{person}{Zhaoqian Xie}, \bibinfo{person}{Hangbo Zhao}, \bibinfo{person}{Mattia Gazzola}, \bibinfo{person}{John~A. Rogers}, {and} \bibinfo{person}{Rashid Bashir}.} \bibinfo{year}{2023}\natexlab{}.
\newblock \showarticletitle{Remote control of muscle-driven miniature robots with battery-free wireless optoelectronics}.
\newblock \bibinfo{journal}{\emph{Science Robotics}} \bibinfo{volume}{8}, \bibinfo{number}{74} (\bibinfo{date}{Jan.} \bibinfo{year}{2023}), \bibinfo{pages}{eadd1053}.
\newblock
\urldef\tempurl%
\url{https://doi.org/10.1126/scirobotics.add1053}
\showDOI{\tempurl}
\newblock
\shownote{Publisher: American Association for the Advancement of Science}.


\bibitem[Kishkinev et~al\mbox{.}(2021)]%
        {kishkinev_navigation_2021}
\bibfield{author}{\bibinfo{person}{Dmitry Kishkinev}, \bibinfo{person}{Florian Packmor}, \bibinfo{person}{Thomas Zechmeister}, \bibinfo{person}{Hans-Christoph Winkler}, \bibinfo{person}{Nikita Chernetsov}, \bibinfo{person}{Henrik Mouritsen}, {and} \bibinfo{person}{Richard~A. Holland}.} \bibinfo{year}{2021}\natexlab{}.
\newblock \showarticletitle{Navigation by extrapolation of geomagnetic cues in a migratory songbird}.
\newblock \bibinfo{journal}{\emph{Current Biology}} \bibinfo{volume}{31}, \bibinfo{number}{7} (\bibinfo{date}{April} \bibinfo{year}{2021}), \bibinfo{pages}{1563--1569.e4}.
\newblock
\showISSN{0960-9822}
\urldef\tempurl%
\url{https://doi.org/10.1016/j.cub.2021.01.051}
\showDOI{\tempurl}
\newblock
\shownote{Publisher: Elsevier}.


\bibitem[Koelle et~al\mbox{.}(2022)]%
        {koelle_prototyping_2022}
\bibfield{author}{\bibinfo{person}{Marion Koelle}, \bibinfo{person}{Madalina Nicolae}, \bibinfo{person}{Aditya~Shekhar Nittala}, \bibinfo{person}{Marc Teyssier}, {and} \bibinfo{person}{Jürgen Steimle}.} \bibinfo{year}{2022}\natexlab{}.
\newblock \showarticletitle{Prototyping {Soft} {Devices} with {Interactive} {Bioplastics}}. In \bibinfo{booktitle}{\emph{Proceedings of the 35th {Annual} {ACM} {Symposium} on {User} {Interface} {Software} and {Technology}}}. \bibinfo{publisher}{ACM}, \bibinfo{address}{Bend OR USA}, \bibinfo{pages}{1--16}.
\newblock
\showISBNx{978-1-4503-9320-1}
\urldef\tempurl%
\url{https://doi.org/10.1145/3526113.3545623}
\showDOI{\tempurl}


\bibitem[Larson et~al\mbox{.}(2011)]%
        {larson_heatwave_2011}
\bibfield{author}{\bibinfo{person}{Eric Larson}, \bibinfo{person}{Gabe Cohn}, \bibinfo{person}{Sidhant Gupta}, \bibinfo{person}{Xiaofeng Ren}, \bibinfo{person}{Beverly Harrison}, \bibinfo{person}{Dieter Fox}, {and} \bibinfo{person}{Shwetak Patel}.} \bibinfo{year}{2011}\natexlab{}.
\newblock \showarticletitle{{HeatWave}: thermal imaging for surface user interaction}. In \bibinfo{booktitle}{\emph{Proceedings of the {SIGCHI} {Conference} on {Human} {Factors} in {Computing} {Systems}}} \emph{(\bibinfo{series}{{CHI} '11})}. \bibinfo{publisher}{Association for Computing Machinery}, \bibinfo{address}{New York, NY, USA}, \bibinfo{pages}{2565--2574}.
\newblock
\showISBNx{978-1-4503-0228-9}
\urldef\tempurl%
\url{https://doi.org/10.1145/1978942.1979317}
\showDOI{\tempurl}


\bibitem[Li et~al\mbox{.}(2012)]%
        {li_determining_2012}
\bibfield{author}{\bibinfo{person}{Chunhui Li}, \bibinfo{person}{Guangying Guan}, \bibinfo{person}{Roberto Reif}, \bibinfo{person}{Zhihong Huang}, {and} \bibinfo{person}{Ruikang~K. Wang}.} \bibinfo{year}{2012}\natexlab{}.
\newblock \showarticletitle{Determining elastic properties of skin by measuring surface waves from an impulse mechanical stimulus using phase-sensitive optical coherence tomography}.
\newblock \bibinfo{journal}{\emph{Journal of the Royal Society Interface}} \bibinfo{volume}{9}, \bibinfo{number}{70} (\bibinfo{date}{May} \bibinfo{year}{2012}), \bibinfo{pages}{831--841}.
\newblock
\showISSN{1742-5689}
\urldef\tempurl%
\url{https://doi.org/10.1098/rsif.2011.0583}
\showDOI{\tempurl}


\bibitem[Li et~al\mbox{.}(2019)]%
        {li_editing_2019}
\bibfield{author}{\bibinfo{person}{Jingyi Li}, \bibinfo{person}{Son Kim}, \bibinfo{person}{Joshua~A. Miele}, \bibinfo{person}{Maneesh Agrawala}, {and} \bibinfo{person}{Sean Follmer}.} \bibinfo{year}{2019}\natexlab{}.
\newblock \showarticletitle{Editing {Spatial} {Layouts} through {Tactile} {Templates} for {People} with {Visual} {Impairments}}. In \bibinfo{booktitle}{\emph{Proceedings of the 2019 {CHI} {Conference} on {Human} {Factors} in {Computing} {Systems}}} \emph{(\bibinfo{series}{{CHI} '19})}. \bibinfo{publisher}{Association for Computing Machinery}, \bibinfo{address}{New York, NY, USA}, \bibinfo{pages}{1--11}.
\newblock
\showISBNx{978-1-4503-5970-2}
\urldef\tempurl%
\url{https://doi.org/10.1145/3290605.3300436}
\showDOI{\tempurl}


\bibitem[Liao et~al\mbox{.}(2023)]%
        {liao_effects_2023}
\bibfield{author}{\bibinfo{person}{Kuotian Liao}, \bibinfo{person}{Paul Grandgeorge}, \bibinfo{person}{Andrew~M. Jimenez}, \bibinfo{person}{Bichlien~H. Nguyen}, {and} \bibinfo{person}{Eleftheria Roumeli}.} \bibinfo{year}{2023}\natexlab{}.
\newblock \showarticletitle{Effects of mechanical cell disruption on the morphology and properties of spirulina-{PLA} biocomposites}.
\newblock \bibinfo{journal}{\emph{Sustainable Materials and Technologies}}  \bibinfo{volume}{36} (\bibinfo{date}{July} \bibinfo{year}{2023}), \bibinfo{pages}{e00591}.
\newblock
\showISSN{2214-9937}
\urldef\tempurl%
\url{https://doi.org/10.1016/j.susmat.2023.e00591}
\showDOI{\tempurl}


\bibitem[Lu et~al\mbox{.}(2020)]%
        {lu_battery-less_2020}
\bibfield{author}{\bibinfo{person}{Haojian Lu}, \bibinfo{person}{Ying Hong}, \bibinfo{person}{Yuanyuan Yang}, \bibinfo{person}{Zhengbao Yang}, {and} \bibinfo{person}{Yajing Shen}.} \bibinfo{year}{2020}\natexlab{}.
\newblock \showarticletitle{Battery-{Less} {Soft} {Millirobot} {That} {Can} {Move}, {Sense}, and {Communicate} {Remotely} by {Coupling} the {Magnetic} and {Piezoelectric} {Effects}}.
\newblock \bibinfo{journal}{\emph{Advanced Science}} \bibinfo{volume}{7}, \bibinfo{number}{13} (\bibinfo{year}{2020}), \bibinfo{pages}{2000069}.
\newblock
\showISSN{2198-3844}
\urldef\tempurl%
\url{https://doi.org/10.1002/advs.202000069}
\showDOI{\tempurl}
\newblock
\shownote{\_eprint: https://onlinelibrary.wiley.com/doi/pdf/10.1002/advs.202000069}.


\bibitem[Melfi et~al\mbox{.}(2020)]%
        {melfi_understanding_2020}
\bibfield{author}{\bibinfo{person}{Giuseppe Melfi}, \bibinfo{person}{Karin Müller}, \bibinfo{person}{Thorsten Schwarz}, \bibinfo{person}{Gerhard Jaworek}, {and} \bibinfo{person}{Rainer Stiefelhagen}.} \bibinfo{year}{2020}\natexlab{}.
\newblock \showarticletitle{Understanding what you feel: {A} {Mobile} {Audio}-{Tactile} {System} for {Graphics} {Used} at {Schools} with {Students} with {Visual} {Impairment}}. In \bibinfo{booktitle}{\emph{Proceedings of the 2020 {CHI} {Conference} on {Human} {Factors} in {Computing} {Systems}}} \emph{(\bibinfo{series}{{CHI} '20})}. \bibinfo{publisher}{Association for Computing Machinery}, \bibinfo{address}{New York, NY, USA}, \bibinfo{pages}{1--12}.
\newblock
\showISBNx{978-1-4503-6708-0}
\urldef\tempurl%
\url{https://doi.org/10.1145/3313831.3376508}
\showDOI{\tempurl}


\bibitem[Meltzoff et~al\mbox{.}(2018)]%
        {meltzoff_infant_2018}
\bibfield{author}{\bibinfo{person}{Andrew~N. Meltzoff}, \bibinfo{person}{Rey~R. Ramírez}, \bibinfo{person}{Joni~N. Saby}, \bibinfo{person}{Eric Larson}, \bibinfo{person}{Samu Taulu}, {and} \bibinfo{person}{Peter~J. Marshall}.} \bibinfo{year}{2018}\natexlab{}.
\newblock \showarticletitle{Infant brain responses to felt and observed touch of hands and feet: an {MEG} study}.
\newblock \bibinfo{journal}{\emph{Developmental Science}} \bibinfo{volume}{21}, \bibinfo{number}{5} (\bibinfo{year}{2018}), \bibinfo{pages}{e12651}.
\newblock
\showISSN{1467-7687}
\urldef\tempurl%
\url{https://doi.org/10.1111/desc.12651}
\showDOI{\tempurl}
\newblock
\shownote{\_eprint: https://onlinelibrary.wiley.com/doi/pdf/10.1111/desc.12651}.


\bibitem[Nicolae et~al\mbox{.}(2023)]%
        {nicolae_biohybrid_2023}
\bibfield{author}{\bibinfo{person}{Madalina Nicolae}, \bibinfo{person}{Vivien Roussel}, \bibinfo{person}{Marion Koelle}, \bibinfo{person}{Samuel Huron}, \bibinfo{person}{Jürgen Steimle}, {and} \bibinfo{person}{Marc Teyssier}.} \bibinfo{year}{2023}\natexlab{}.
\newblock \showarticletitle{Biohybrid {Devices}: {Prototyping} {Interactive} {Devices} with {Growable} {Materials}}. In \bibinfo{booktitle}{\emph{Proceedings of the 36th {Annual} {ACM} {Symposium} on {User} {Interface} {Software} and {Technology}}} \emph{(\bibinfo{series}{{UIST} '23})}. \bibinfo{publisher}{Association for Computing Machinery}, \bibinfo{address}{New York, NY, USA}, \bibinfo{pages}{1--15}.
\newblock
\showISBNx{9798400701320}
\urldef\tempurl%
\url{https://doi.org/10.1145/3586183.3606774}
\showDOI{\tempurl}


\bibitem[Nisser et~al\mbox{.}(2022)]%
        {nisser_mixels_2022}
\bibfield{author}{\bibinfo{person}{Martin Nisser}, \bibinfo{person}{Yashaswini Makaram}, \bibinfo{person}{Lucian Covarrubias}, \bibinfo{person}{Amadou~Yaye Bah}, \bibinfo{person}{Faraz Faruqi}, \bibinfo{person}{Ryo Suzuki}, {and} \bibinfo{person}{Stefanie Mueller}.} \bibinfo{year}{2022}\natexlab{}.
\newblock \showarticletitle{Mixels: {Fabricating} {Interfaces} using {Programmable} {Magnetic} {Pixels}}. In \bibinfo{booktitle}{\emph{Proceedings of the 35th {Annual} {ACM} {Symposium} on {User} {Interface} {Software} and {Technology}}} \emph{(\bibinfo{series}{{UIST} '22})}. \bibinfo{publisher}{Association for Computing Machinery}, \bibinfo{address}{New York, NY, USA}, \bibinfo{pages}{1--12}.
\newblock
\showISBNx{978-1-4503-9320-1}
\urldef\tempurl%
\url{https://doi.org/10.1145/3526113.3545698}
\showDOI{\tempurl}


\bibitem[Ogunseitan et~al\mbox{.}(2022)]%
        {ogunseitan_biobased_2022}
\bibfield{author}{\bibinfo{person}{Oladele~A. Ogunseitan}, \bibinfo{person}{Julie~M. Schoenung}, \bibinfo{person}{Johnny Lincoln}, \bibinfo{person}{Bichlien~H. Nguyen}, \bibinfo{person}{Karin Strauss}, \bibinfo{person}{Kali Frost}, \bibinfo{person}{Eric Schwartz}, \bibinfo{person}{Haoyang He}, {and} \bibinfo{person}{Maryam Ibrahim}.} \bibinfo{year}{2022}\natexlab{}.
\newblock \showarticletitle{Biobased materials for sustainable printed circuit boards}.
\newblock \bibinfo{journal}{\emph{Nature Reviews Materials}} \bibinfo{volume}{7}, \bibinfo{number}{10} (\bibinfo{date}{Oct.} \bibinfo{year}{2022}), \bibinfo{pages}{749--750}.
\newblock
\showISSN{2058-8437}
\urldef\tempurl%
\url{https://doi.org/10.1038/s41578-022-00485-2}
\showDOI{\tempurl}
\newblock
\shownote{Publisher: Nature Publishing Group}.


\bibitem[Oprişan et~al\mbox{.}(2016)]%
        {oprisan_experimental_2016}
\bibfield{author}{\bibinfo{person}{C Oprişan}, \bibinfo{person}{V Cârlescu}, \bibinfo{person}{A Barnea}, \bibinfo{person}{Gh Prisacaru}, \bibinfo{person}{D~N Olaru}, {and} \bibinfo{person}{Gh Plesu}.} \bibinfo{year}{2016}\natexlab{}.
\newblock \showarticletitle{Experimental determination of the {Young}'s modulus for the fingers with application in prehension systems for small cylindrical objects}.
\newblock \bibinfo{journal}{\emph{IOP Conference Series: Materials Science and Engineering}}  \bibinfo{volume}{147} (\bibinfo{date}{Aug.} \bibinfo{year}{2016}), \bibinfo{pages}{012058}.
\newblock
\showISSN{1757-8981, 1757-899X}
\urldef\tempurl%
\url{https://doi.org/10.1088/1757-899X/147/1/012058}
\showDOI{\tempurl}


\bibitem[Parks and Akhtar(1968)]%
        {parks_magnetic_1968}
\bibfield{author}{\bibinfo{person}{George~A. Parks} {and} \bibinfo{person}{Salim Akhtar}.} \bibinfo{year}{1968}\natexlab{}.
\newblock \showarticletitle{Magnetic moment of {Fe2}+ in paramagnetic minerals}.
\newblock \bibinfo{journal}{\emph{American Mineralogist}} \bibinfo{volume}{53}, \bibinfo{number}{3-4} (\bibinfo{date}{April} \bibinfo{year}{1968}), \bibinfo{pages}{406--415}.
\newblock
\showISSN{0003-004X}


\bibitem[Pawlaczyk et~al\mbox{.}(2013)]%
        {pawlaczyk_age-dependent_2013}
\bibfield{author}{\bibinfo{person}{Mariola Pawlaczyk}, \bibinfo{person}{Monika Lelonkiewicz}, {and} \bibinfo{person}{Michał Wieczorowski}.} \bibinfo{year}{2013}\natexlab{}.
\newblock \showarticletitle{Age-dependent biomechanical properties of the skin}.
\newblock \bibinfo{journal}{\emph{Advances in Dermatology and Allergology/Postȩpy Dermatologii i Alergologii}} \bibinfo{volume}{30}, \bibinfo{number}{5} (\bibinfo{date}{Oct.} \bibinfo{year}{2013}), \bibinfo{pages}{302--306}.
\newblock
\showISSN{1642-395X}
\urldef\tempurl%
\url{https://doi.org/10.5114/pdia.2013.38359}
\showDOI{\tempurl}


\bibitem[Phutane et~al\mbox{.}(2022)]%
        {phutane_tactile_2022}
\bibfield{author}{\bibinfo{person}{Mahika Phutane}, \bibinfo{person}{Julie Wright}, \bibinfo{person}{Brenda~Veronica Castro}, \bibinfo{person}{Lei Shi}, \bibinfo{person}{Simone~R. Stern}, \bibinfo{person}{Holly~M. Lawson}, {and} \bibinfo{person}{Shiri Azenkot}.} \bibinfo{year}{2022}\natexlab{}.
\newblock \showarticletitle{Tactile {Materials} in {Practice}: {Understanding} the {Experiences} of {Teachers} of the {Visually} {Impaired}}.
\newblock \bibinfo{journal}{\emph{ACM Transactions on Accessible Computing}} \bibinfo{volume}{15}, \bibinfo{number}{3} (\bibinfo{date}{July} \bibinfo{year}{2022}), \bibinfo{pages}{17:1--17:34}.
\newblock
\showISSN{1936-7228}
\urldef\tempurl%
\url{https://doi.org/10.1145/3508364}
\showDOI{\tempurl}


\bibitem[Rolsky and Kelkar(2021)]%
        {rolsky_degradation_2021}
\bibfield{author}{\bibinfo{person}{Charles Rolsky} {and} \bibinfo{person}{Varun Kelkar}.} \bibinfo{year}{2021}\natexlab{}.
\newblock \showarticletitle{Degradation of {Polyvinyl} {Alcohol} in {US} {Wastewater} {Treatment} {Plants} and {Subsequent} {Nationwide} {Emission} {Estimate}}.
\newblock \bibinfo{journal}{\emph{International Journal of Environmental Research and Public Health}} \bibinfo{volume}{18}, \bibinfo{number}{11} (\bibinfo{date}{Jan.} \bibinfo{year}{2021}), \bibinfo{pages}{6027}.
\newblock
\showISSN{1660-4601}
\urldef\tempurl%
\url{https://doi.org/10.3390/ijerph18116027}
\showDOI{\tempurl}
\newblock
\shownote{Number: 11 Publisher: Multidisciplinary Digital Publishing Institute}.


\bibitem[Sato et~al\mbox{.}(2012)]%
        {sato_touche_2012}
\bibfield{author}{\bibinfo{person}{Munehiko Sato}, \bibinfo{person}{Ivan Poupyrev}, {and} \bibinfo{person}{Chris Harrison}.} \bibinfo{year}{2012}\natexlab{}.
\newblock \showarticletitle{Touché: enhancing touch interaction on humans, screens, liquids, and everyday objects}. In \bibinfo{booktitle}{\emph{Proceedings of the {SIGCHI} {Conference} on {Human} {Factors} in {Computing} {Systems}}}. \bibinfo{publisher}{ACM}, \bibinfo{address}{Austin Texas USA}, \bibinfo{pages}{483--492}.
\newblock
\showISBNx{978-1-4503-1015-4}
\urldef\tempurl%
\url{https://doi.org/10.1145/2207676.2207743}
\showDOI{\tempurl}


\bibitem[Tee et~al\mbox{.}(2015)]%
        {tee_skin-inspired_2015}
\bibfield{author}{\bibinfo{person}{Benjamin C.-K. Tee}, \bibinfo{person}{Alex Chortos}, \bibinfo{person}{Andre Berndt}, \bibinfo{person}{Amanda~Kim Nguyen}, \bibinfo{person}{Ariane Tom}, \bibinfo{person}{Allister McGuire}, \bibinfo{person}{Ziliang~Carter Lin}, \bibinfo{person}{Kevin Tien}, \bibinfo{person}{Won-Gyu Bae}, \bibinfo{person}{Huiliang Wang}, \bibinfo{person}{Ping Mei}, \bibinfo{person}{Ho-Hsiu Chou}, \bibinfo{person}{Bianxiao Cui}, \bibinfo{person}{Karl Deisseroth}, \bibinfo{person}{Tse~Nga Ng}, {and} \bibinfo{person}{Zhenan Bao}.} \bibinfo{year}{2015}\natexlab{}.
\newblock \showarticletitle{A skin-inspired organic digital mechanoreceptor}.
\newblock \bibinfo{journal}{\emph{Science}} \bibinfo{volume}{350}, \bibinfo{number}{6258} (\bibinfo{date}{Oct.} \bibinfo{year}{2015}), \bibinfo{pages}{313--316}.
\newblock
\urldef\tempurl%
\url{https://doi.org/10.1126/science.aaa9306}
\showDOI{\tempurl}
\newblock
\shownote{Publisher: American Association for the Advancement of Science}.


\bibitem[Tsai et~al\mbox{.}(2019)]%
        {tsai_dielectrical_2019}
\bibfield{author}{\bibinfo{person}{B. Tsai}, \bibinfo{person}{H. Xue}, \bibinfo{person}{E. Birgersson}, \bibinfo{person}{S. Ollmar}, {and} \bibinfo{person}{U. Birgersson}.} \bibinfo{year}{2019}\natexlab{}.
\newblock \showarticletitle{Dielectrical {Properties} of {Living} {Epidermis} and {Dermis} in the {Frequency} {Range} from 1 {kHz} to 1 {MHz}}.
\newblock \bibinfo{journal}{\emph{Journal of Electrical Bioimpedance}} \bibinfo{volume}{10}, \bibinfo{number}{1} (\bibinfo{date}{July} \bibinfo{year}{2019}), \bibinfo{pages}{14--23}.
\newblock
\showISSN{1891-5469}
\urldef\tempurl%
\url{https://doi.org/10.2478/joeb-2019-0003}
\showDOI{\tempurl}


\bibitem[Vasquez and Vega(2019)]%
        {vasquez_myco-accessories_2019}
\bibfield{author}{\bibinfo{person}{Eldy S.~Lazaro Vasquez} {and} \bibinfo{person}{Katia Vega}.} \bibinfo{year}{2019}\natexlab{}.
\newblock \showarticletitle{Myco-accessories: sustainable wearables with biodegradable materials}. In \bibinfo{booktitle}{\emph{Proceedings of the 2019 {ACM} {International} {Symposium} on {Wearable} {Computers}}} \emph{(\bibinfo{series}{{ISWC} '19})}. \bibinfo{publisher}{Association for Computing Machinery}, \bibinfo{address}{New York, NY, USA}, \bibinfo{pages}{306--311}.
\newblock
\showISBNx{978-1-4503-6870-4}
\urldef\tempurl%
\url{https://doi.org/10.1145/3341163.3346938}
\showDOI{\tempurl}


\bibitem[Waghmare et~al\mbox{.}(2023)]%
        {waghmare_z-ring_2023}
\bibfield{author}{\bibinfo{person}{Anandghan Waghmare}, \bibinfo{person}{Youssef Ben~Taleb}, \bibinfo{person}{Ishan Chatterjee}, \bibinfo{person}{Arjun Narendra}, {and} \bibinfo{person}{Shwetak Patel}.} \bibinfo{year}{2023}\natexlab{}.
\newblock \showarticletitle{Z-{Ring}: {Single}-{Point} {Bio}-{Impedance} {Sensing} for {Gesture}, {Touch}, {Object} and {User} {Recognition}}. In \bibinfo{booktitle}{\emph{Proceedings of the 2023 {CHI} {Conference} on {Human} {Factors} in {Computing} {Systems}}} \emph{(\bibinfo{series}{{CHI} '23})}. \bibinfo{publisher}{Association for Computing Machinery}, \bibinfo{address}{New York, NY, USA}, \bibinfo{pages}{1--18}.
\newblock
\showISBNx{978-1-4503-9421-5}
\urldef\tempurl%
\url{https://doi.org/10.1145/3544548.3581422}
\showDOI{\tempurl}


\bibitem[Waghmare et~al\mbox{.}(2020)]%
        {waghmare_ubiquitouch_2020}
\bibfield{author}{\bibinfo{person}{Anandghan Waghmare}, \bibinfo{person}{Qiuyue Xue}, \bibinfo{person}{Dingtian Zhang}, \bibinfo{person}{Yuhui Zhao}, \bibinfo{person}{Shivan Mittal}, \bibinfo{person}{Nivedita Arora}, \bibinfo{person}{Ceara Byrne}, \bibinfo{person}{Thad Starner}, {and} \bibinfo{person}{Gregory~D Abowd}.} \bibinfo{year}{2020}\natexlab{}.
\newblock \showarticletitle{{UbiquiTouch}: {Self} {Sustaining} {Ubiquitous} {Touch} {Interfaces}}.
\newblock \bibinfo{journal}{\emph{Proceedings of the ACM on Interactive, Mobile, Wearable and Ubiquitous Technologies}} \bibinfo{volume}{4}, \bibinfo{number}{1} (\bibinfo{date}{March} \bibinfo{year}{2020}), \bibinfo{pages}{27:1--27:22}.
\newblock
\urldef\tempurl%
\url{https://doi.org/10.1145/3380989}
\showDOI{\tempurl}


\bibitem[Wang et~al\mbox{.}(2023)]%
        {wang_neuromorphic_2023}
\bibfield{author}{\bibinfo{person}{Weichen Wang}, \bibinfo{person}{Yuanwen Jiang}, \bibinfo{person}{Donglai Zhong}, \bibinfo{person}{Zhitao Zhang}, \bibinfo{person}{Snehashis Choudhury}, \bibinfo{person}{Jian-Cheng Lai}, \bibinfo{person}{Huaxin Gong}, \bibinfo{person}{Simiao Niu}, \bibinfo{person}{Xuzhou Yan}, \bibinfo{person}{Yu Zheng}, \bibinfo{person}{Chien-Chung Shih}, \bibinfo{person}{Rui Ning}, \bibinfo{person}{Qing Lin}, \bibinfo{person}{Deling Li}, \bibinfo{person}{Yun-Hi Kim}, \bibinfo{person}{Jingwan Kim}, \bibinfo{person}{Yi-Xuan Wang}, \bibinfo{person}{Chuanzhen Zhao}, \bibinfo{person}{Chengyi Xu}, \bibinfo{person}{Xiaozhou Ji}, \bibinfo{person}{Yuya Nishio}, \bibinfo{person}{Hao Lyu}, \bibinfo{person}{Jeffrey B.-H. Tok}, {and} \bibinfo{person}{Zhenan Bao}.} \bibinfo{year}{2023}\natexlab{}.
\newblock \showarticletitle{Neuromorphic sensorimotor loop embodied by monolithically integrated, low-voltage, soft e-skin}.
\newblock \bibinfo{journal}{\emph{Science}} \bibinfo{volume}{380}, \bibinfo{number}{6646} (\bibinfo{date}{May} \bibinfo{year}{2023}), \bibinfo{pages}{735--742}.
\newblock
\urldef\tempurl%
\url{https://doi.org/10.1126/science.ade0086}
\showDOI{\tempurl}
\newblock
\shownote{Publisher: American Association for the Advancement of Science}.


\bibitem[Wang et~al\mbox{.}(2022)]%
        {wang_all-biomass-based_2022}
\bibfield{author}{\bibinfo{person}{Yuyuan Wang}, \bibinfo{person}{Xiaoqian Zhang}, \bibinfo{person}{Lijun Kan}, \bibinfo{person}{Feng Shen}, \bibinfo{person}{Hao Ling}, {and} \bibinfo{person}{Xiaohui Wang}.} \bibinfo{year}{2022}\natexlab{}.
\newblock \showarticletitle{All-biomass-based eco-friendly waterproof coating for paper-based green packaging}.
\newblock \bibinfo{journal}{\emph{Green Chemistry}} \bibinfo{volume}{24}, \bibinfo{number}{18} (\bibinfo{date}{Sept.} \bibinfo{year}{2022}), \bibinfo{pages}{7039--7048}.
\newblock
\showISSN{1463-9270}
\urldef\tempurl%
\url{https://doi.org/10.1039/D2GC02265F}
\showDOI{\tempurl}
\newblock
\shownote{Publisher: The Royal Society of Chemistry}.


\bibitem[Wexler(1966)]%
        {wexler_size_1966}
\bibfield{author}{\bibinfo{person}{G. Wexler}.} \bibinfo{year}{1966}\natexlab{}.
\newblock \showarticletitle{The size effect and the non-local {Boltzmann} transport equation in orifice and disk geometry}.
\newblock \bibinfo{journal}{\emph{Proceedings of the Physical Society}} \bibinfo{volume}{89}, \bibinfo{number}{4} (\bibinfo{date}{Dec.} \bibinfo{year}{1966}), \bibinfo{pages}{927}.
\newblock
\showISSN{0370-1328}
\urldef\tempurl%
\url{https://doi.org/10.1088/0370-1328/89/4/316}
\showDOI{\tempurl}


\bibitem[Wilson(2010)]%
        {wilson_using_2010}
\bibfield{author}{\bibinfo{person}{Andrew~D. Wilson}.} \bibinfo{year}{2010}\natexlab{}.
\newblock \showarticletitle{Using a depth camera as a touch sensor}. In \bibinfo{booktitle}{\emph{{ACM} {International} {Conference} on {Interactive} {Tabletops} and {Surfaces}}} \emph{(\bibinfo{series}{{ITS} '10})}. \bibinfo{publisher}{Association for Computing Machinery}, \bibinfo{address}{New York, NY, USA}, \bibinfo{pages}{69--72}.
\newblock
\showISBNx{978-1-4503-0399-6}
\urldef\tempurl%
\url{https://doi.org/10.1145/1936652.1936665}
\showDOI{\tempurl}


\bibitem[Xiao et~al\mbox{.}(2016)]%
        {xiao_direct_2016}
\bibfield{author}{\bibinfo{person}{Robert Xiao}, \bibinfo{person}{Scott Hudson}, {and} \bibinfo{person}{Chris Harrison}.} \bibinfo{year}{2016}\natexlab{}.
\newblock \showarticletitle{{DIRECT}: {Making} {Touch} {Tracking} on {Ordinary} {Surfaces} {Practical} with {Hybrid} {Depth}-{Infrared} {Sensing}}. In \bibinfo{booktitle}{\emph{Proceedings of the 2016 {ACM} {International} {Conference} on {Interactive} {Surfaces} and {Spaces}}} \emph{(\bibinfo{series}{{ISS} '16})}. \bibinfo{publisher}{Association for Computing Machinery}, \bibinfo{address}{New York, NY, USA}, \bibinfo{pages}{85--94}.
\newblock
\showISBNx{978-1-4503-4248-3}
\urldef\tempurl%
\url{https://doi.org/10.1145/2992154.2992173}
\showDOI{\tempurl}


\bibitem[Zhang et~al\mbox{.}(2022)]%
        {zhang_flexible_2022}
\bibfield{author}{\bibinfo{person}{Dingtian Zhang}, \bibinfo{person}{Canek Fuentes-Hernandez}, \bibinfo{person}{Raaghesh Vijayan}, \bibinfo{person}{Yang Zhang}, \bibinfo{person}{Yunzhi Li}, \bibinfo{person}{Jung~Wook Park}, \bibinfo{person}{Yiyang Wang}, \bibinfo{person}{Yuhui Zhao}, \bibinfo{person}{Nivedita Arora}, \bibinfo{person}{Ali Mirzazadeh}, \bibinfo{person}{Youngwook Do}, \bibinfo{person}{Tingyu Cheng}, \bibinfo{person}{Saiganesh Swaminathan}, \bibinfo{person}{Thad Starner}, \bibinfo{person}{Trisha~L. Andrew}, {and} \bibinfo{person}{Gregory~D. Abowd}.} \bibinfo{year}{2022}\natexlab{}.
\newblock \showarticletitle{Flexible computational photodetectors for self-powered activity sensing}.
\newblock \bibinfo{journal}{\emph{npj Flexible Electronics}} \bibinfo{volume}{6}, \bibinfo{number}{1} (\bibinfo{date}{Jan.} \bibinfo{year}{2022}), \bibinfo{pages}{1--8}.
\newblock
\showISSN{2397-4621}
\urldef\tempurl%
\url{https://doi.org/10.1038/s41528-022-00137-z}
\showDOI{\tempurl}
\newblock
\shownote{Publisher: Nature Publishing Group}.


\bibitem[Zhang et~al\mbox{.}(2020)]%
        {zhang_optosense_2020}
\bibfield{author}{\bibinfo{person}{Dingtian Zhang}, \bibinfo{person}{Jung~Wook Park}, \bibinfo{person}{Yang Zhang}, \bibinfo{person}{Yuhui Zhao}, \bibinfo{person}{Yiyang Wang}, \bibinfo{person}{Yunzhi Li}, \bibinfo{person}{Tanvi Bhagwat}, \bibinfo{person}{Wen-Fang Chou}, \bibinfo{person}{Xiaojia Jia}, \bibinfo{person}{Bernard Kippelen}, \bibinfo{person}{Canek Fuentes-Hernandez}, \bibinfo{person}{Thad Starner}, {and} \bibinfo{person}{Gregory~D. Abowd}.} \bibinfo{year}{2020}\natexlab{}.
\newblock \showarticletitle{{OptoSense}: {Towards} {Ubiquitous} {Self}-{Powered} {Ambient} {Light} {Sensing} {Surfaces}}.
\newblock \bibinfo{journal}{\emph{Proceedings of the ACM on Interactive, Mobile, Wearable and Ubiquitous Technologies}} \bibinfo{volume}{4}, \bibinfo{number}{3} (\bibinfo{date}{Sept.} \bibinfo{year}{2020}), \bibinfo{pages}{103:1--103:27}.
\newblock
\urldef\tempurl%
\url{https://doi.org/10.1145/3411826}
\showDOI{\tempurl}


\bibitem[Zhang et~al\mbox{.}(2018)]%
        {zhang_interactiles_2018}
\bibfield{author}{\bibinfo{person}{Xiaoyi Zhang}, \bibinfo{person}{Tracy Tran}, \bibinfo{person}{Yuqian Sun}, \bibinfo{person}{Ian Culhane}, \bibinfo{person}{Shobhit Jain}, \bibinfo{person}{James Fogarty}, {and} \bibinfo{person}{Jennifer Mankoff}.} \bibinfo{year}{2018}\natexlab{}.
\newblock \showarticletitle{Interactiles: {3D} {Printed} {Tactile} {Interfaces} to {Enhance} {Mobile} {Touchscreen} {Accessibility}}. In \bibinfo{booktitle}{\emph{Proceedings of the 20th {International} {ACM} {SIGACCESS} {Conference} on {Computers} and {Accessibility}}} \emph{(\bibinfo{series}{{ASSETS} '18})}. \bibinfo{publisher}{Association for Computing Machinery}, \bibinfo{address}{New York, NY, USA}, \bibinfo{pages}{131--142}.
\newblock
\showISBNx{978-1-4503-5650-3}
\urldef\tempurl%
\url{https://doi.org/10.1145/3234695.3236349}
\showDOI{\tempurl}


\bibitem[Zhang et~al\mbox{.}(2016)]%
        {zhang_advancing_2016}
\bibfield{author}{\bibinfo{person}{Yang Zhang}, \bibinfo{person}{Robert Xiao}, {and} \bibinfo{person}{Chris Harrison}.} \bibinfo{year}{2016}\natexlab{}.
\newblock \showarticletitle{Advancing {Hand} {Gesture} {Recognition} with {High} {Resolution} {Electrical} {Impedance} {Tomography}}. In \bibinfo{booktitle}{\emph{Proceedings of the 29th {Annual} {Symposium} on {User} {Interface} {Software} and {Technology}}} \emph{(\bibinfo{series}{{UIST} '16})}. \bibinfo{publisher}{Association for Computing Machinery}, \bibinfo{address}{New York, NY, USA}, \bibinfo{pages}{843--850}.
\newblock
\showISBNx{978-1-4503-4189-9}
\urldef\tempurl%
\url{https://doi.org/10.1145/2984511.2984574}
\showDOI{\tempurl}


\bibitem[Zhang et~al\mbox{.}(2023)]%
        {zhang_recyclable_2023}
\bibfield{author}{\bibinfo{person}{Zhihan Zhang}, \bibinfo{person}{Agni~K. Biswal}, \bibinfo{person}{Ankush Nandi}, \bibinfo{person}{Kali Frost}, \bibinfo{person}{Jake~A. Smith}, \bibinfo{person}{Bichlien~H. Nguyen}, \bibinfo{person}{Shwetak Patel}, \bibinfo{person}{Aniruddh Vashisth}, {and} \bibinfo{person}{Vikram Iyer}.} \bibinfo{year}{2023}\natexlab{}.
\newblock \bibinfo{title}{Recyclable vitrimer-based printed circuit board for circular electronics}.
\newblock
\newblock
\urldef\tempurl%
\url{https://doi.org/10.48550/arXiv.2308.12496}
\showDOI{\tempurl}
\newblock
\shownote{arXiv:2308.12496 [physics]}.


\bibitem[Zhang et~al\mbox{.}(2024)]%
        {zhang_deltalca_2024}
\bibfield{author}{\bibinfo{person}{Zhihan Zhang}, \bibinfo{person}{Felix Hähnlein}, \bibinfo{person}{Yuxuan Mei}, \bibinfo{person}{Zachary Englhardt}, \bibinfo{person}{Shwetak Patel}, \bibinfo{person}{Adriana Schulz}, {and} \bibinfo{person}{Vikram Iyer}.} \bibinfo{year}{2024}\natexlab{}.
\newblock \showarticletitle{{DeltaLCA}: {Comparative} {Life}-{Cycle} {Assessment} for {Electronics} {Design}}.
\newblock \bibinfo{journal}{\emph{Proceedings of the ACM on Interactive, Mobile, Wearable and Ubiquitous Technologies}} \bibinfo{volume}{8}, \bibinfo{number}{1} (\bibinfo{date}{March} \bibinfo{year}{2024}), \bibinfo{pages}{29:1--29:29}.
\newblock
\urldef\tempurl%
\url{https://doi.org/10.1145/3643561}
\showDOI{\tempurl}


\end{thebibliography}

\appendix
\clearpage
\section{Appendix}

\subsection{Detailed Theory for Electrical Encoding}
\subsubsection{Contact Impedance} \label{Detailed_ContactImpedance}
Our theoretical framework made the following simplifying assumptions:
\begin{enumerate}
    \item Contact surfaces are clean without insulating barriers, as well as continuous and frictionless.
    \item There is no tensile stress at the contact point.
    \item User's finger is a perfectly elastic solid and skin surface is smooth, i.e., no friction ridges and lines; external interactive material surface is a perfectly rigid body.
\end{enumerate}

Under these conditions, $Z_{contact}$ is governed by the resistivity, and the size of the contact radius relative to the electron mean free path, which transitions between Sharvin and diffuse scattering mechanism, as explained in the Maxwell spreading resistance formula \cite{wexler_size_1966}:
\begin{equation}
    Z_{contact} = \frac{4\rho^{*} l_{e}}{3\pi a^{2}} + v\left(l_{e}/a\right)\frac{\rho^{*}}{2a}
\label{eq:Wexler}
\end{equation}
where $\rho$ is the resistivity separating two bodies, $l_{e}$ is the electron mean free path length of the material, $a$ is the contact radius, $v$ is a function of the ratio $l_{e}/a$, and $\rho^{*}$ is $\frac{\rho_1+\rho_2}{2}$, $\rho_1, \rho_2$ are the resistivity on two sides of the contact, respectively. Given the dominance of the diffuse scattering mechanism in our context $a \gg l_{e}$, so that
\begin{align}
    \lim\limits_{{l_e}/a \to 0} v\left({l_e}/a\right) &= 1 \nonumber \\
    \frac{4\rho^{*} l_{e}}{3\pi a^{2}} &\approx 0 \nonumber 
\label{eq:le/a}
\end{align}
Now Eq.\ref{eq:Wexler} can be simplified to
\begin{equation}
    Z_{contact} = 0 + 1\times\frac{\rho^{*}}{2a} = \frac{\rho^{*}}{2a} \Leftrightarrow \frac{\rho_{finger}+\rho_{material}}{4a}
\label{eq:Wexler_simplified}
\end{equation}

\paragraph{Hertzian Contact Stress Model} This relationship is further refined through the Hertzian contact stress model, accounting for the applied force and the elastic properties of the finger and interactive material, to express the effective contact radius of the finger pad $a$ and, consequently, the contact area $A$. For contact radius $a$, we have
\begin{equation}
    a = \left[ \frac{3FR^{*}}{4E^{*}} \right]^{\frac{1}{3}}
\label{eq:Hertzian}
\end{equation}
where $F$ is the force applied over the contact area of the finger and material surface, $R^{*}$ is the effective radius of curvature of the finger and material surface, and $E^{*}$ is the effective modulus of elasticity of the system. $R^{*}$ is defined as
\begin{equation}
    \frac{1}{R^{*}} = \frac{1}{R_{finger}}+\frac{1}{R_{material}} 
\label{eq:effective_radius}
\end{equation}

Since we assume the external interactive material surface is a perfectly rigid body, for a rigid surface, its radius of curvature $R_{material}$ can be considered infinitely large:
\begin{align}
    \lim\limits_{R_{material} \to \infty} \left(\frac{1}{R_{finger}}+\frac{1}{R_{material}} \right) &= \frac{1}{R_{finger}} \nonumber \\
    \Rightarrow \frac{1}{R^{*}} = \frac{1}{R_{finger}} \Leftrightarrow R^{*} &= R_{finger}
\label{eq:effective_radius_2}
\end{align}

By incorporating the mechanics of contact between disparate materials, specifically between a sphere (i.e., the user's finger) and a flat surface (i.e., the material). We define the effective modulus of elasticity $E^{*}$ as a function of the individual modulus of elasticity $E$ and Poisson ratios $v$ of the interacting bodies. This relationship is quantified as follows
\begin{equation}
    \frac{1}{E^{*}} = \frac{3}{4} \times \left( \frac{1-{v_{finger}}^{2}}{E_{finger}}+\frac{1-{v_{material}}^{2}}{E_{material}} \right)
\label{eq:elasticity}
\end{equation}

Given the substantially higher modulus of elasticity for rigid object surfaces compared to elastic biological tissues, $E_{material} \gg E_{finger}$, therefore Eq.\ref{eq:elasticity} can be simplified with the material's contribution becoming negligible:
\begin{equation}
    \Rightarrow \frac{1}{E^{*}} \approx \frac{3}{4} \times \frac{1-{v_{finger}}^{2}}{E_{finger}}
\label{eq:elasticity_2}
\end{equation}

This leads us to a simplified Hertzian contact stress model from Eq. \ref{eq:Hertzian}, which allows us to calculate the contact radius $a$ considering the force applied during the touch interaction and the inherent physical properties of the finger:
\begin{equation}
    a = \left[ \frac{9F R_{finger}} {16 \left( \frac{E_{finger}}{1-{v_{finger}}^{2}} \right)} \right]^{\frac{1}{3}}
\label{eq:Hertzian_final}
\end{equation}

In validating our theoretical contact stress model of the finger and interactive material, empirical results from recent studies characterizing human skin tissue and finger mechanics are used: a Poisson ratio for human skin tissue of 0.48 \cite{li_determining_2012}; an average tap force of 0.50 N with a standard deviation of 0.09 N \cite{asakawa_fingertip_2017}; the elasticity of the finger, as characterized by Young’s modulus, presents a wide range due to factors such as applied normal load, gender differences, and the anatomical variance between the skin's superficial and deeper layers adjacent to the bone, fluctuating between 0.2 MPa \cite{oprisan_experimental_2016} and 4.6 MPa \cite{pawlaczyk_age-dependent_2013}. For the purposes of our analysis, the effective curvature radius of the finger is approximated as 1 cm. Substituting these empirically derived values into our final Hertzian contact stress model Eq.\ref{eq:Hertzian_final}, we derive the effective contact radius: 
\begin{align}
    a &\in [0.84 mm, 2.40 mm]
\label{eq:radius_number}
\end{align}

These results align with empirical observations---19.9 and 14.0 mm for the orientations of 30° and 45° \cite{dzidek_contact_2017}. This stress model is a critical component in understanding the contact interface of tactile interactions. A study also characterized the resistivity of the living epidermis (the outermost layer of skin) of fingers to be 11 $\Omega \text{m}$ at 1 MHz \cite{tsai_dielectrical_2019}. By integrating these datums into Eq.\ref{eq:Wexler_simplified}, we derive the contact impedance, $Z_{contact}$, as a function of material resistivity:

\begin{equation}
    Z_{contact} = \frac{11 \Omega \text{m}+\rho_{material}}{4a}, a \in [0.84 mm, 2.40 mm]
\label{eq:Wexler_final_Appendix}
\end{equation}

\subsubsection{Interactive Material Impedance} \label{Detailed_MaterialImpedance}

The intrinsic impedance $Z_{material}$ diverges in behavior under alternating current (AC) conditions, necessitating a distinction from direct current (DC) scenarios where impedance equates to resistance, $R$. In the AC paradigm, where common capacitive sensors and bio-impedance sensors operate at, $Z_{material}$ embodies both resistive and reactive components:
\begin{equation}
    Z_{material} = R+jX
\label{eq:Z_material}
\end{equation}
where $j$ is the imaginary unit and $X = \omega L - \frac{1}{\omega C}$, $\omega$ is the angular frequency $2\pi f$. Assume interactive materials as regular solid cuboids with constant area $A$ and thickness $t$ for simplification, we deduce expressions for resistance, $R = \rho \frac{t}{A}$, and capacitance, $C = \frac{t}{\omega \epsilon A}$. Due to the diamagnetism of common conductive materials such as copper, silver, and graphite, which results in low magnetic permeability, the inductance part could be negligible. Now we have Eq.\ref{eq:Z_material} as
\begin{equation}
    Z_{material} = \rho \frac{t}{A} - j\frac{t}{\omega \epsilon A} = \frac{t}{A} (\rho - j\frac{1}{\omega \epsilon})
\label{eq:Z_material_2}
\end{equation}
where $\epsilon$ is the permittivity of IM.

We know that in the theory of wave causality in classical electrodynamics, permittivity is a complex quantity and can be expressed as $\epsilon = \epsilon' + j\epsilon'' = \epsilon_0 + j \frac{\sigma}{\omega } \Rightarrow j\frac{1}{\omega \epsilon} = j\frac{1}{\omega  (\epsilon_0 + j \frac{\sigma}{\omega })} = \frac{-\omega }{j\omega^2 \epsilon_0 - \sigma}$, where $\epsilon_0$ is the vacuum permittivity. Thus, Eq.\ref{eq:Z_material_2} arrives at:

\begin{equation}
    Z_{material} = \frac{t}{A} (\frac{1}{\sigma} + \frac{\omega }{j\omega^2 \epsilon_0 - \sigma})
\label{eq:Z_material_3}
\end{equation}

\subsection{Material Characterization}
\subsubsection{Conductivity} \label{Conductivity}
The conductivity tests were conducted using a sourcemeter (Keithley 2470). IM beams with dimensions $55\times8\times1.5 \text{ mm}$ were prepared as test specimens. The conductivity is measured by connecting a source meter test lead to each end of the long side of the specimen. The current is measured while applying 40 volts DC across the specimen. The ambient temperature is maintained at 25 °C during the interval that electrical measurements are being made. The conductivity of IM is calculated by the following formula: $\sigma = \frac{1}{\rho} = \frac{L}{RA}$, where R is the measured resistance, A is the effective cross-section area, and L is the length of the specimen.

\subsubsection{Flexural Strength} \label{Flexural}

Flexural strength test was conducted on a Shimadzu AGS-X mechanical test frame. Biodegradable IMs with dimensions $55\times8\times2 \text{ mm}$ were prepared as test specimens. Specimens are centered on the span supports with the long axis perpendicular to the crosshead. The test is performed under ambient conditions, and the load is applied at a constant rate of crosshead movement of 0.5\% per second. The flexural strength is calculated using the following formula: $S = (3PL)/(2Wt)$, where P is the loading at breaking, L is the span, and W, t are the width, thickness of the specimen, respectively. 

\subsubsection{Magnetization} \label{Magnetization}
Magnetization studies were performed using a Quantum Design MPMS3 SQUID magnetometer. 2 mm thick samples are cut to lateral dimensions of approximately 4 × 4 mm to fit within a drinking straw for low-background sample mounting. Two smaller straw pieces are inserted at each end of the straw to hold the sample in place from above and below. For all measurements, the magnetic field is applied perpendicular to the sample and swept in a full hysteresis loop between ± 3 T. Within ± 0.3 T, we measure using 100 Oe intervals and 500 Oe intervals outside of this range. Each measurement uses a DC scan length of 3cm and a scan time of 4 s. Calibration of the magnetic field is performed by subjecting a 0.2582 g Pd standard to the same sequence of fields at a temperature of 298 K, where the magnetic susceptibility of Pd is known to be $5\times10^{-6}$ emu/Oe-g. Reported field strengths are then divided by the moment to yield calibrated field values.

\subsubsection{Contact Angle} \label{ContactAngle}
Contact angle tests were performed using a Krüss Drop Shape Analyzer to evaluate the surface energy of our IMs. For each processing condition, five separate droplets of 4 ± 1 $\mu$L of deionized water are placed on the surface of the specimen. The contact angle is then determined as the average of angles formed between the sample surface and the left and right tangent (ellipse fit) of the droplet edge.

\subsection{Supplementary Figures and Tables}

\begin{figure}[h]
  \centering
  \includegraphics[width=0.8\linewidth]{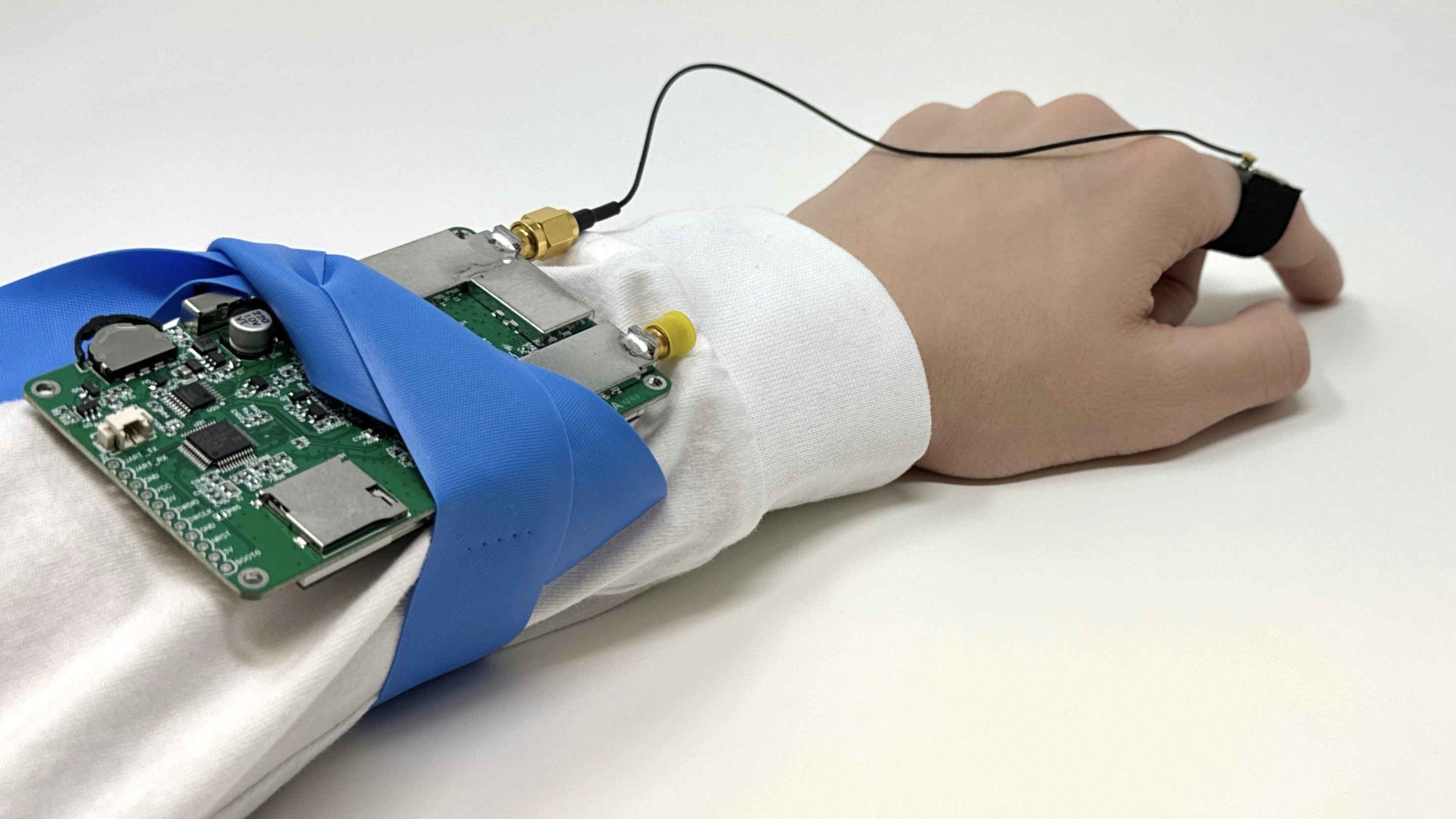}
  \includegraphics[width=0.8\linewidth]{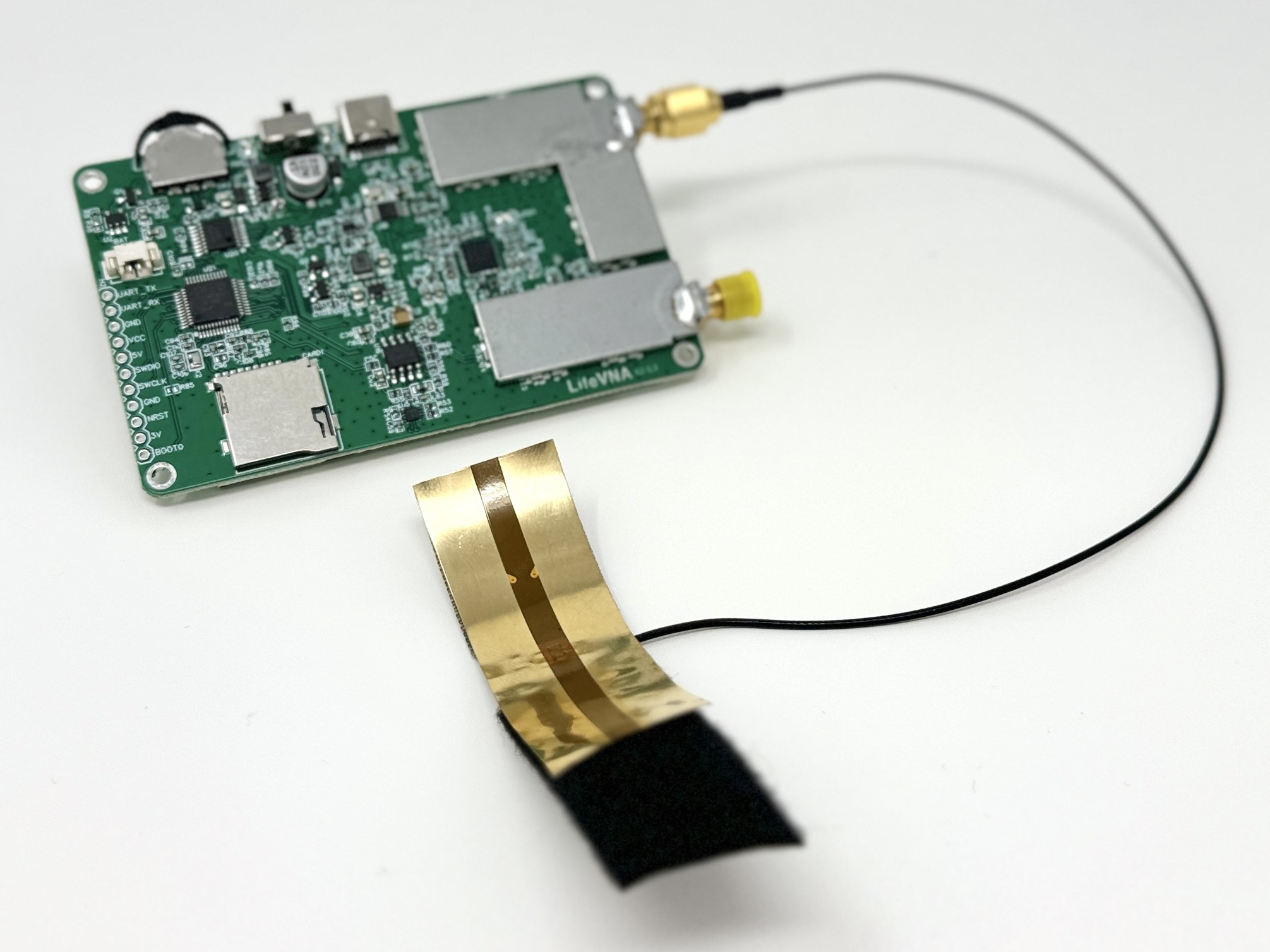}
    \caption{Single-point bio-impedance sensing setup. A LiteVNA connected to a flexible PCB with exposed copper electrodes. Hook and loop fasteners are glued to the backside of each end of the flexible PCB to allow the electrodes to fit conformally around the user’s finger. The electrodes are positioned on the proximal phalange of the user's contact finger.}
    \label{fig:bioimpedance_setup}
\end{figure}

\begin{figure}[h]
  \centering
  \includegraphics[width=\linewidth]{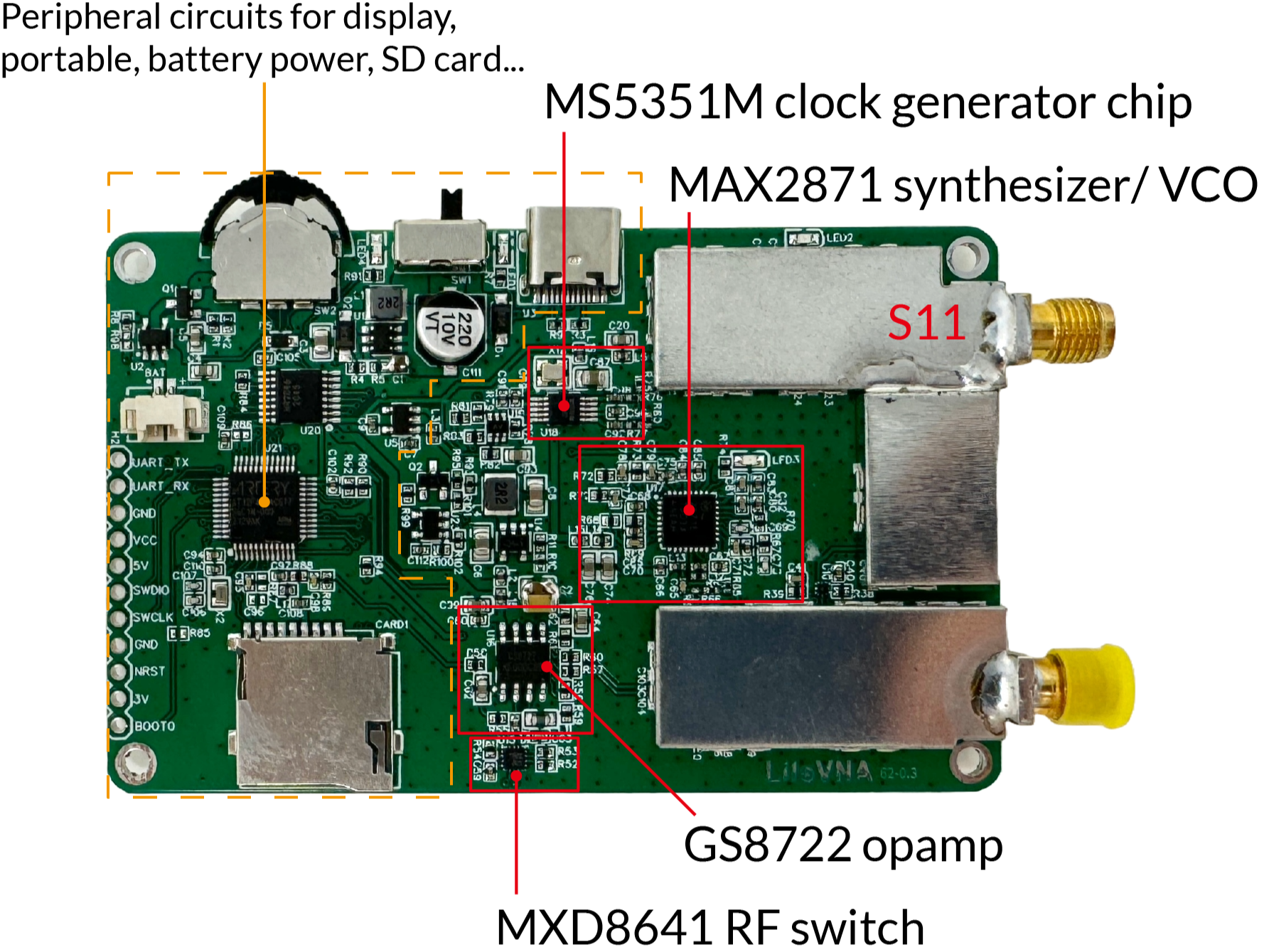}
    \caption{Teardown of our bio-impedance sensor. According to a rough estimate, only 15\% of the area of the commercial sensor is the actual circuitry required for bio-impedance measurements; the rest is peripheral circuitry for display, battery power and so on.}
    \label{fig:VNA_teardiwn}
\end{figure}

\begin{figure}[h]
  \centering
  \includegraphics[width=\linewidth]{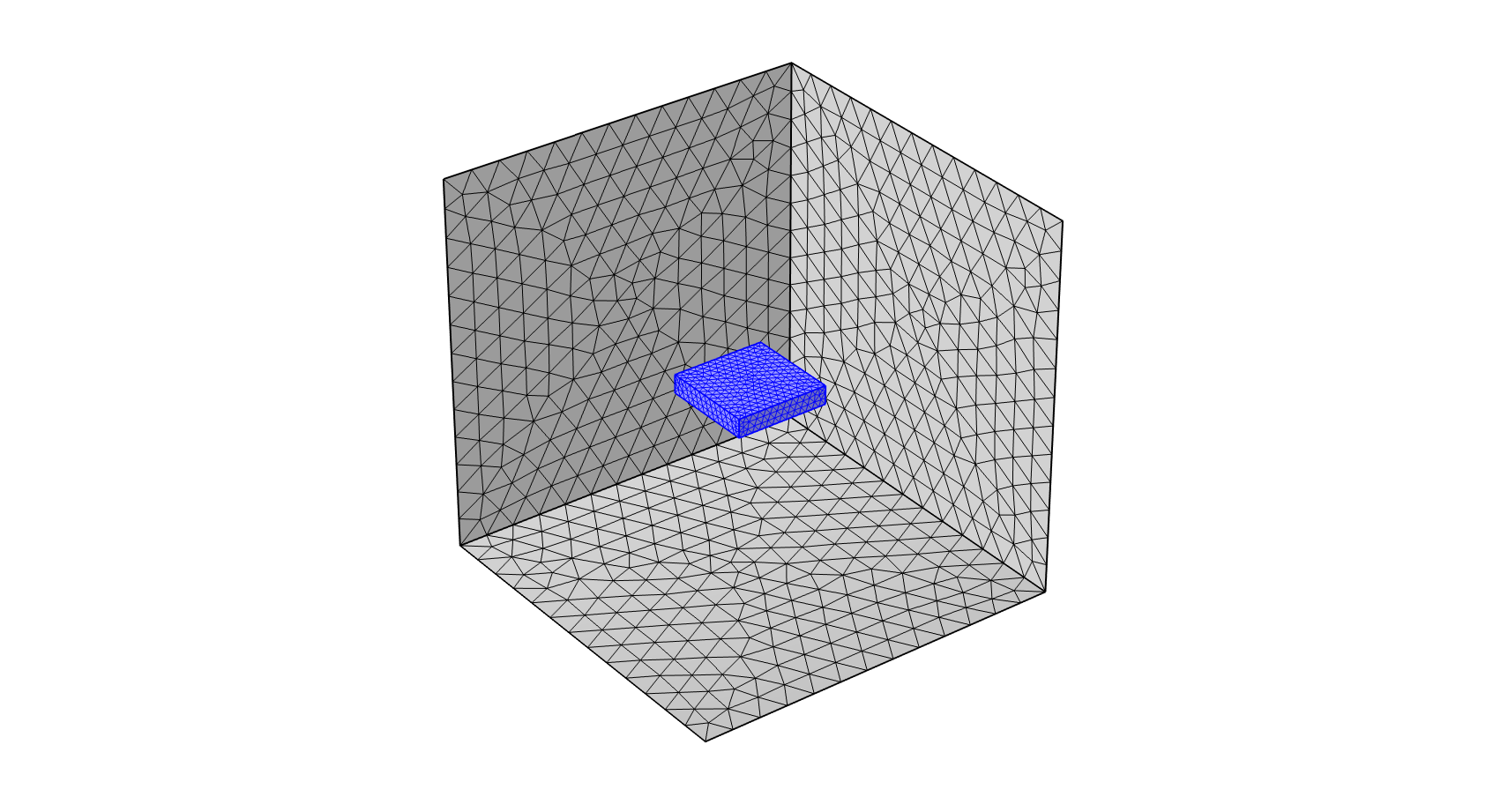}
    \caption{COMSOL magnetic simulation setup. A rectangular cuboid with dimensions of $20\times20\times5 \text{mm}$ representing the magnetic IM is placed at the spatial center of a $100\times100\times100 \text{mm}$ air region.}
    \label{fig:comsol_setup}
\end{figure}

\begin{figure}[h]
  \centering
  \includegraphics[width=\linewidth]{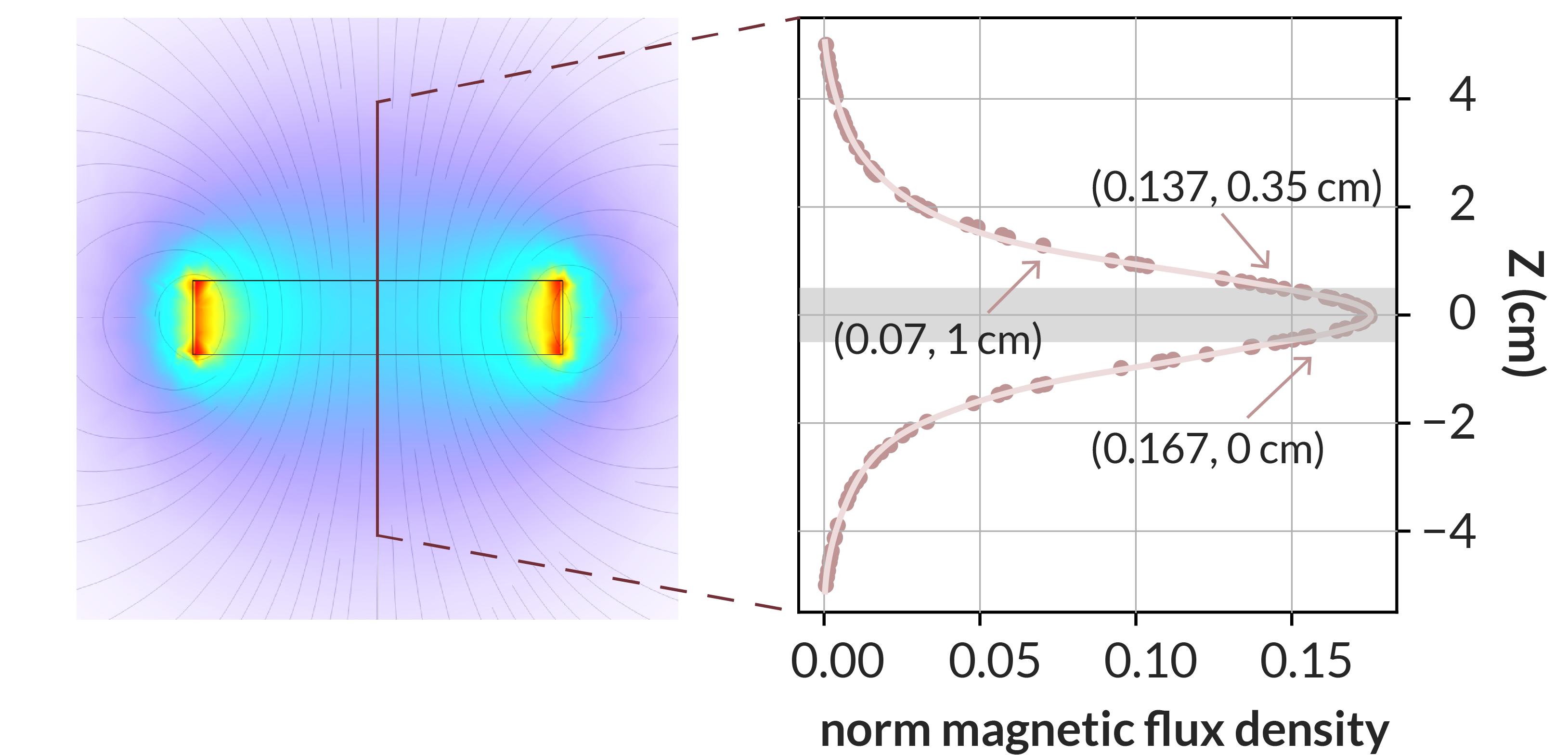}
    \caption{Simulated magnetic flux density along the Z-axis perpendicular to the center top of the magnetic IM, gray shaded region indicates the magnetic field inside the magnetic IM.}
    \label{fig:magnetism_comsol_z}
\end{figure}


\begin{figure}[h]
  \centering
  \includegraphics[width=\linewidth]{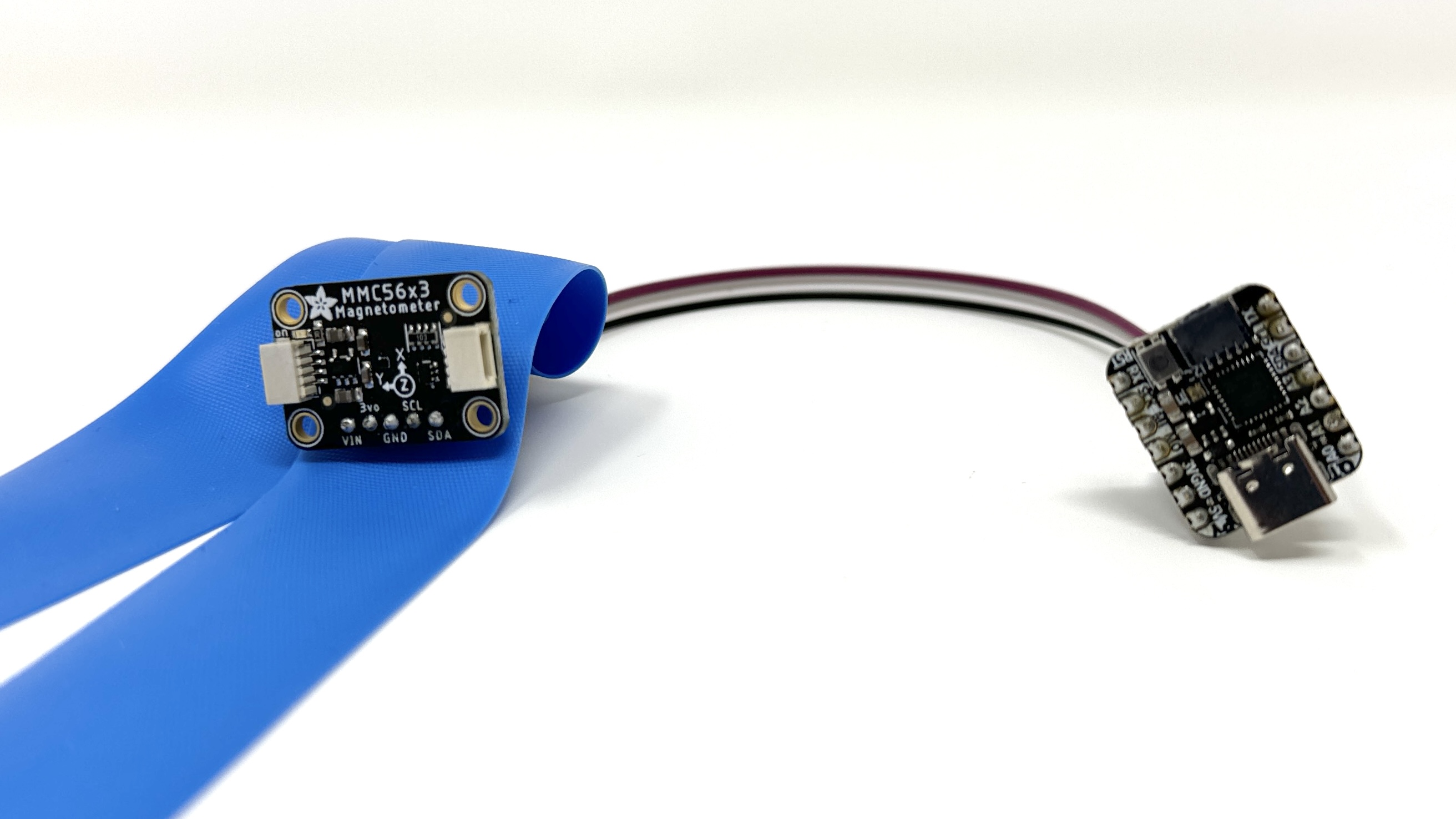}
    \caption{Magnetic sensing wearable. A magnetometer Memsic MMC5603 is connected to a QT Py SAMD21. An elastic bandage is used to secure the magnetometer on the user's olecranon and an MCU to the cubital fossa.}
    \label{fig:magnetometer_MCU_setup}
\end{figure}

\begin{table*}[b]
  \centering
  \includegraphics[width=\linewidth]{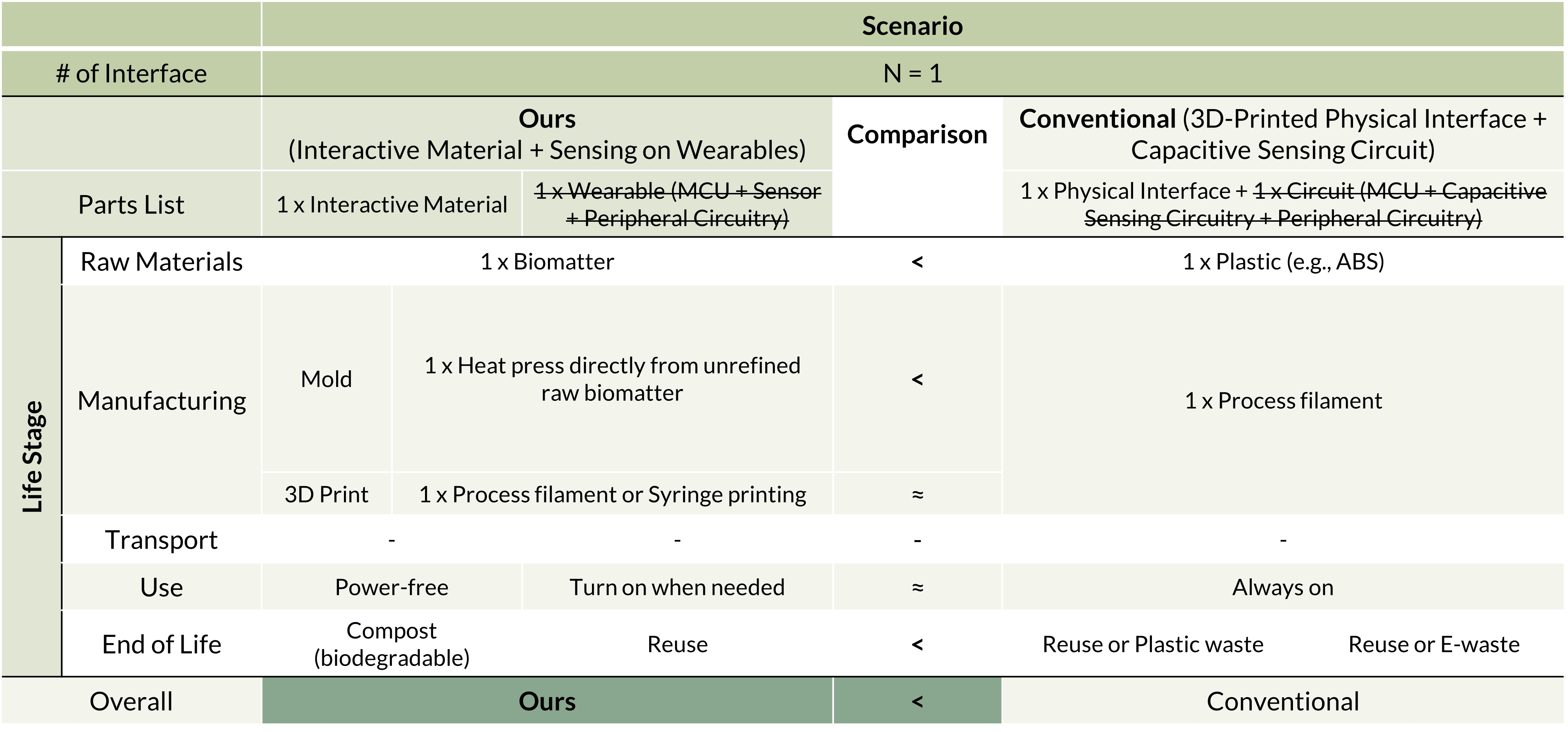}
  \includegraphics[width=\linewidth]{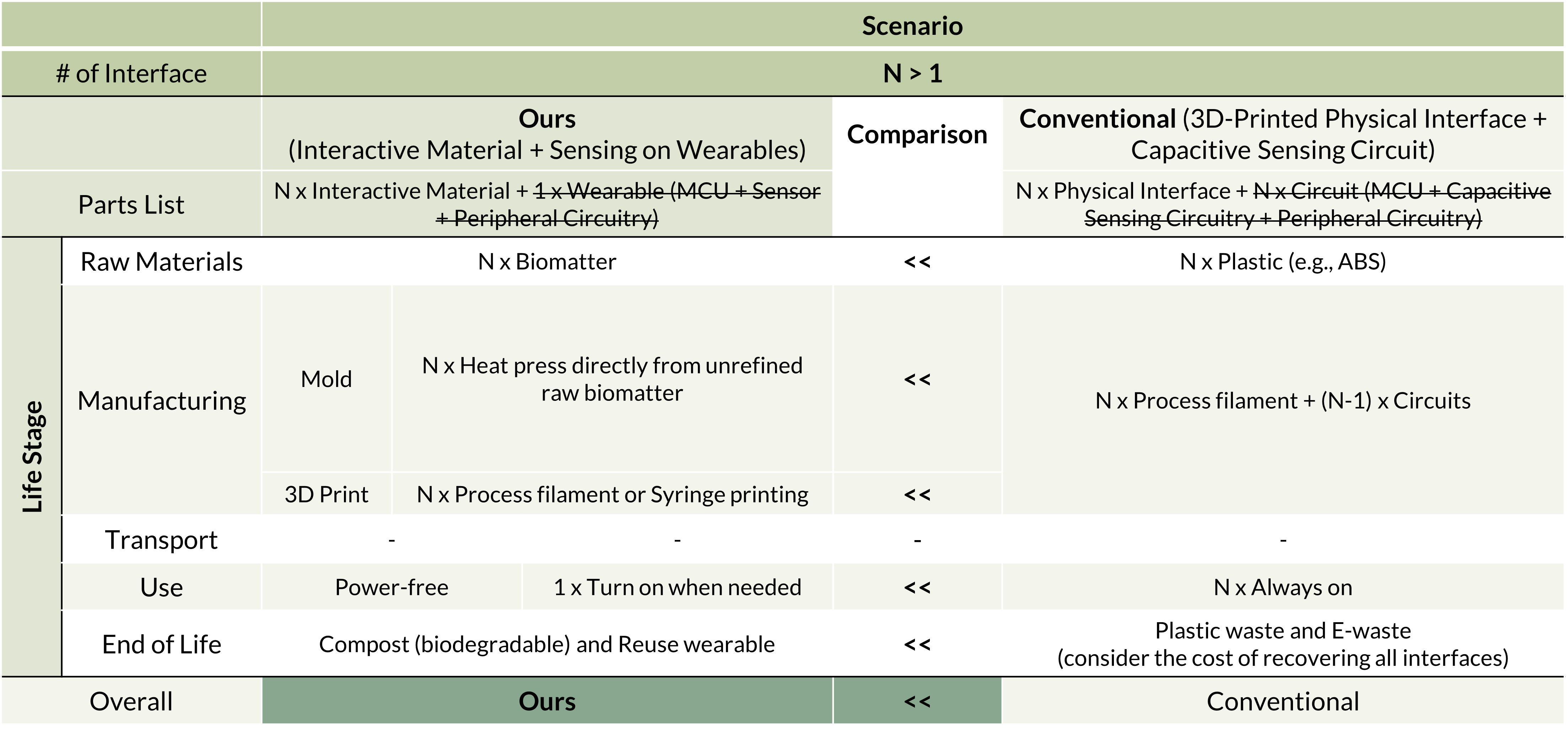}
    \caption{Detailed cradle-to-grave environmental impact comparison of Biodegradable Interactive Materials vs. conventional approach in single (top) and multiple (bottom) interactive interfaces scenario.
    }
    \label{tab:detailed_LCA}
\end{table*}

\begin{table*}[b]
  \centering
  \begin{tabular}{c|c|c|c|c|c|c}
    \toprule
    Time & Temperature (°F) & Dew Point (°F) & Humidity (\%) & Wind Speed (mph) & Pressure (in) & Precipitation (in) \\
    \midrule
    Day & Avg & Avg & Avg & Avg & Avg & Total \\
    \midrule
    0 & 41.5 & 35.3 & 79.7 & 4.2 & 29.7 & 0 \\
    1 & 42.3 & 32.5 & 69.2 & 4.3 & 29.7 & 0 \\
    2 & 46.8 & 41.8 & 82.9 & 10.5 & 29.6 & 0 \\
    3 & 45 & 42.3 & 90.5 & 7.8 & 29.6 & 0.34 \\
    4 & 43.6 & 37.6 & 79.8 & 8 & 29.6 & 0.26 \\
    5 & 41.6 & 27.3 & 60.8 & 11.8 & 29.4 & 0 \\
    6 & 39.4 & 30.4 & 70.9 & 8.9 & 29.5 & 0.17 \\
    7 & 43 & 17.6 & 37.2 & 10.6 & 29.7 & 0.11 \\
    8 & 46.5 & 19.8 & 34.7 & 12.5 & 29.5 & 0 \\
    9 & 45.9 & 35 & 67.2 & 5.9 & 29.4 & 0 \\
    10 & 43.9 & 38.2 & 80.6 & 6.2 & 29.3 & 0 \\
    11 & 47.2 & 42.4 & 83.7 & 4.8 & 29.4 & 0.07 \\
    12 & 47.4 & 42.6 & 83.3 & 7.9 & 29.5 & 0.18 \\
    13 & 48.6 & 38.8 & 70.2 & 5.1 & 29.8 & 0.33 \\
    14 & 40.9 & 37.5 & 88.2 & 3.2 & 29.9 & 0 \\
    15 & 44.9 & 40.8 & 85.8 & 7.6 & 29.7 & 0 \\
    16 & 45.8 & 40.6 & 82.5 & 16.4 & 29.3 & 0 \\
    17 & 37.2 & 29.8 & 75 & 9 & 29.4 & 0.17 \\
    18 & 39.1 & 32.1 & 76.4 & 9.5 & 29.5 & 0.05 \\
    19 & 46.8 & 41.6 & 82.3 & 15.8 & 29.1 & 0.1 \\
    20 & 37.6 & 33.9 & 86.7 & 10.7 & 29 & 1.01 \\
    21 & 38.1 & 32.5 & 80.7 & 8.8 & 29 & 0.49 \\
    \bottomrule
  \end{tabular}
  \caption{Daily weather data during the biodegradation test.}
  \label{tab:biodegradation_daily_weather_data}
\end{table*}

\clearpage

\end{document}